\documentclass[1p,11pt,compress]{elsarticle}

\usepackage{amsmath,amssymb,amsfonts}
\usepackage{bm}
\usepackage{graphicx}
\usepackage{booktabs}
\usepackage[colorlinks=true,linkcolor=blue,urlcolor=blue,citecolor=blue]{hyperref}
\usepackage{libertine}
\usepackage{pgfplots}
\pgfplotsset{compat=1.16}
\usetikzlibrary{arrows.meta}

\newcommand{\Tr}{\mathrm{Tr}}
\newcommand{\e}{\varepsilon}
\newcommand{\Rey}{\mathrm{Re}}
\newcommand{\Imy}{\mathrm{Im}}

\usepackage{xcolor}

\journal{Annals of Physics}

\begin{document}

\begin{frontmatter}

\title{Kinetic Lifshitz invariants and dynamics of nonreciprocal fluctuations in superconductors}

\author[uw]{Tony Liu}
\author[cbpf,uw]{Joaquim Telles de Miranda}
\author[uw]{Daniel Shaffer}
\author[uw]{Alex Levchenko}

\address[uw]{Department of Physics, University of Wisconsin--Madison, Madison, Wisconsin 53706, USA}
\address[cbpf]{Centro Brasileiro de Pesquisas F\'isicas (CBPF), Rio de Janeiro, Brazil}

\begin{abstract}
We derive the generalized time-dependent Ginzburg-Landau theory of a disordered noncentrosymmetric superconductor from the Keldysh nonlinear sigma model, using a two-dimensional electron gas with Rashba spin-orbit coupling and an in-plane Zeeman field as a minimal model. On the thermodynamic side we construct the Lifshitz invariants of the free energy, the linear and cubic gradient terms and the momentum-odd part of the quartic vertex, and trace their dependence on disorder. The couplings of the quadratic sector are governed by a single closed-form kernel controlled by the ratio of the Dyakonov-Perel spin-relaxation rate to temperature. It interpolates between the weak-relaxation regime, where the invariants are suppressed, and the relaxation-dominated regime, where the helical modulation of the order parameter saturates at a universal, disorder-independent value. This crossover reconciles known results for the magnetoelectric couplings of disordered Rashba superconductors. Because the theory is formulated on the Keldysh contour, it also determines the dissipative dynamics: the relaxation rate of a fluctuation with pair momentum $\bm q$, and hence the Langevin noise power, acquires a term odd in $\bm q$ and odd in the magnetic field. The structure of this kinetic Lifshitz invariant is dictated by Onsager reciprocity: friction and noise change together, so that equal-time fluctuations remain Gibbsian while the relaxation dynamics becomes nonreciprocal. As applications we compute the superconducting diode efficiency near $T_c$, where the cubic invariant competes with the quartic vertex and with higher-gradient terms, reversing the sign of the diode coefficient. We also consider fluctuation-induced magnetochiral anisotropy above $T_c$, which grows as the inverse square of the distance to $T_c$ and is suppressed in the relaxation-dominated regime.
\end{abstract}


\end{frontmatter}

\newpage

\tableofcontents

%==============================================================================================================================================================================================
%==============================================================================================================================================================================================
%==============================================================================================================================================================================================
%==============================================================================================================================================================================================
%==============================================================================================================================================================================================
\section{Introduction}
\label{sec:intro}

In the Landau theory of second-order phase transitions \cite{Landau:1937}, Lifshitz invariants \cite{Lifshitz:1941} are symmetry-allowed gradient terms in the free energy expansion that couple an order parameter to its spatial derivatives. A Lifshitz invariant destabilizes the spatially uniform phase and leads to a modulated, incommensurate, or helical phase at the transition. The crucial symmetry requirement for a Lifshitz invariant to exist is the absence of inversion symmetry (spatial parity $\mathcal{P}$). Indeed, mathematically, a Lifshitz invariant involves two components of a multicomponent order parameter (or an order parameter and its conjugate), say $\eta_i$ and $\eta_j$, and takes the general form
\begin{equation}
L_{ijk} = \eta_i \frac{\partial \eta_j}{\partial x_k} - \eta_j \frac{\partial \eta_i}{\partial x_k}.
\end{equation}
Because the spatial derivative flips sign under the inversion, but the order parameter product does not (whether the components are both even, both odd, or complex conjugates of one another), the entire Lifshitz term changes sign under inversion, $L_{ijk} \to -L_{ijk}$. Since the free energy density must be a scalar that is strictly invariant under all symmetry operations of the high-temperature (parent) phase, such a term cannot exist in a centrosymmetric system. If, however, the parent phase is chiral or otherwise lacks inversion symmetry, the Lifshitz invariant is permitted.

Lifshitz invariants are ubiquitous across physics. In magnetic systems the order parameter is the magnetization vector $\bm{M}(\bm{r})$, and in noncentrosymmetric crystal lattices the lack of inversion symmetry allows for a magnetic Lifshitz invariant known as the Dzyaloshinskii-Moriya interaction \cite{Dzyaloshinsky:1958,Moriya:1960}, with free energy density $F_{\rm L} = D\,\bm{M}\cdot[\bm{\nabla}\times\bm{M}]$. Instead of a simple ferromagnetic alignment, this term forces the spins to cant continuously, giving rise to helical magnetic phases and to topologically protected vortex-like spin textures like magnetic skyrmions \cite{Bogdanov:1989,Muhlbauer:2009}. In chiral nematic liquid crystals the Frank-Oseen elastic energy contains the analogous invariant $F_{\rm L}=\kappa\,\bm{n}\cdot[\bm{\nabla}\times\bm{n}]$ for the director field $\bm n$, which twists the uniform nematic into the cholesteric helix \cite{deGennes:1993}. The flexoelectric effect couples polarization to strain gradients, $F_{\rm L}=-f_{ijkl}P_i\,\partial_l\varepsilon_{jk}$, inducing modulated ferroelectric textures such as polar vortices \cite{Zubko:2013}. The list of examples can be continued.

Central to our study is the case of a noncentrosymmetric superconductor \cite{BauerSigrist:2012,Smidman:2017}, where the order parameter is a complex scalar field, denoted $\Delta(\bm r)$ throughout this paper. In the presence of a magnetic field $\bm{B}$ the gradient term is built from the covariant derivative $\bm{\partial}=-i\bm{\nabla}-2e\bm{A}$, and the Lifshitz invariant takes the form $F_{\rm L}=K_{ij}B_i\,[\Delta^*\partial_j\Delta+\Delta(\partial_j\Delta)^*]$. In the model studied below the field enters only through the Zeeman energy $\bm h$ of the electron spins; for a two-dimensional film in an in-plane field this is the only coupling, and we use $\bm h$ rather than $\bm B$ from Sec.~\ref{sec:overview} on. This coupling breaks the degeneracy between Cooper pairs moving in opposite directions and shifts the minimum of the free energy to a finite pair momentum: the phase of the order parameter winds in space and the system condenses into a helical superconducting state \cite{MineevSamokhin:1994,Agterberg:2003,Barzykin:2002,Kaur:2005,DimitrovaFeigelman:2007}. Note that breaking time-reversal symmetry $\mathcal{T}$ (here by $\bm B$ or by an exchange field) is the second crucial ingredient: for a single complex order parameter the parity-odd bilinear $\Delta^*\partial_j\Delta+{\rm c.c.}$ is a total derivative unless multiplied by a $\mathcal T$-odd prefactor.

At the level of terms linear in spatial gradients, Lifshitz invariants have been classified for all noncentrosymmetric point groups \cite{Samokhin:2004,MineevSamokhin:2008}. Beyond inducing the helical phase they produce a family of magnetoelectric phenomena: (i) an unconventional Meissner response in which the screened field rotates inside the superconductor \cite{Yip:2002,BauerSigrist:2012}; (ii) the anomalous Josephson effect, in which the current-phase relation is shifted by a phase $\varphi_0$ proportional to the Lifshitz coupling \cite{Buzdin:2008,Konschelle:2015,Assouline:2019}; (iii) in-plane-field oscillations of the critical current of wide junctions, governed by the helical wave vector rather than by the flux through the junction area \cite{Kaur:2005,DimitrovaFeigelman:2007}; (iv) the spin polarization induced by a supercurrent \cite{Edelstein:1989,Edelstein:1995,GorkovRashba:2001}.

Higher-order Lifshitz invariants with three or more odd numbers of gradients, or momentum-odd corrections to the quartic vertex, have been studied far less extensively. One might expect such contributions to be small within the gradient expansion and therefore unimportant. This is not so, for a simple reason: the leading (linear) invariant can be removed from the quadratic form of the free energy by a uniform shift of the pair momentum, so several observables of current interest are controlled entirely by the subleading invariants. The most prominent example is the superconducting diode effect (SDE), the nonreciprocity of the critical current with respect to the current direction \cite{Ando:2020,Baumgartner:2022,Nadeem:2023,Shaffer:2025}: the efficiency of this effect is determined by a combination of the cubic gradient term, the momentum-odd part of the quartic term \cite{Daido:2022,YuanFu:2022,HeTanakaLaw:2022,IlicBergeret:2022,Hasan:2025}, and, less obviously, the ordinary even higher-gradient terms, which acquire an odd component once the expansion is referenced to the helical ground state \cite{Hasan:2024,Hasan:2025}. A second example is electrical magnetochiral anisotropy (MCA): a correction to the resistance that is linear in both the bias current and the magnetic field \cite{Rikken:2001,Rikken:2005,TokuraNagaosa:2018}. Experiments indicate that superconductors exhibiting the SDE also display a strongly enhanced MCA near the onset of the superconducting transition \cite{Wakatsuki:2017,Itahashi:2020}, which points to superconducting fluctuations as the amplifying mechanism \cite{WakatsukiNagaosa:2018,Hoshino:2018,DeMiranda:2026,JTM:2026}.

While the general form of all these invariants follows from symmetry alone, their coefficients and their dependence on disorder must be derived microscopically. With few exceptions \cite{DimitrovaFeigelman:2007,HouzetMeyer:2015,IlicBergeret:2022,Daido:2022,VBT:2022,Hasan:2024,IlicBergeret:2024, Hasan:2025, NunchotYanase:2026}, such calculations remain scarce, and the results reported for the paradigmatic disordered Rashba model by different methods have appeared to be in mutual conflict. Resolving this issue is the first goal of the present paper: we show in Sec.~\ref{sec:GL} that the seemingly contradictory answers are the two limits of a single closed-form crossover function of the parameter $\Gamma_t/4\pi T$, where $\Gamma_t=2(\alpha_R p_F)^2\tau$ is the Dyakonov-Perel (DP) spin-relaxation rate.

Our second goal is dictated by the method. We employ the Keldysh nonlinear sigma model (NLSM) of disordered superconductors \cite{FLS:2000,KamenevAndreev:1999,KamenevBook,LevchenkoKamenev:2007,Liao:2017}, generalized to spin-orbit-coupled systems following Refs.~\cite{BergeretTokatly:2014,VBT:2021,VBT:2022}. Unlike approaches based on the free energy or on the Matsubara diagram technique, the Keldysh formulation yields the complete dynamical theory of the order parameter: the time-dependent Ginzburg-Landau (TDGL) equation together with its stochastic Langevin forces \cite{LevchenkoKamenev:2007}. This allows us to pose a question that, to our knowledge, has not been addressed before: do Lifshitz-type nonreciprocal couplings appear in the quantum (Keldysh) sector of the effective action, that is, in the dissipation and in the noise? We show that they do, compute them, and demonstrate that their structure is fixed by two general principles: the Onsager reciprocity of kinetic coefficients and the fluctuation-dissipation theorem (FDT). The resulting stochastic TDGL theory of a noncentrosymmetric superconductor, Eqs.~\eqref{eq:TDGL} and \eqref{eq:noisefinal} below, is the main result of this work.

The paper is organized as follows. Section~\ref{sec:overview} summarizes the main results and the underlying physical picture in nontechnical terms. Section~\ref{sec:formalism} formulates the Keldysh sigma model for the Rashba superconductor and derives the Gaussian action of the superconducting (cooperon) modes, with emphasis on the physical meaning of each coupling. Section~\ref{sec:GL} performs the remaining energy integrals in closed form, presents the generalized Ginzburg-Landau functional, and analyzes the disorder crossover of the Lifshitz invariants and of the helical state. Section~\ref{sec:TDGL} is devoted to the dynamical sector: the nonreciprocal friction, the Onsager analysis, the generalized FDT, and the stochastic TDGL equation. Section~\ref{sec:apps} applies the theory to the diode effect below $T_c$ and to the fluctuation-induced magnetochiral anisotropy above $T_c$. We conclude in Sec.~\ref{sec:conclusions}. Four appendices contain the technical material: the cooperon expansion and the verification of all matrix traces (\ref{app:gaussian}), the master energy integrals and the crossover kernel (\ref{app:integrals}), the quartic vertex (\ref{app:quartic}), and the nonlinear fluctuation transport calculation (\ref{app:MCA}). We set $\hbar=k_B=1$ throughout.

%==============================================================================================================================================================================================
%==============================================================================================================================================================================================
%==============================================================================================================================================================================================
%==============================================================================================================================================================================================
%==============================================================================================================================================================================================
\section{Physical picture and summary of the main results}
\label{sec:overview}

Before entering the formalism we explain, in plain terms, where the various effects come from and how they are organized. The reader interested only in the results may treat this section, together with Table~\ref{tab:couplings}, as a self-contained summary.

\subsection{Gauge covariance of spin-orbit coupling}

For a linear-in-momentum spin-orbit coupling, such as the Rashba interaction, the kinetic Hamiltonian can be written as $(\bm p+\bm{\mathcal A})^2/2m$ with an SU(2)-matrix-valued vector potential $\bm{\mathcal A}$, while the Zeeman field plays the role of the temporal component $\mathcal A_0$ \cite{BergeretTokatly:2014}. This observation has an important consequence: to linear order the SU(2) potential can be removed by a local gauge rotation of the electron spin, so no physical magnetoelectric effect can appear at first order in Rashba coupling $\alpha_R$. Observable effects must be built from the gauge-invariant field strength
\begin{equation}
F_{\mu\nu}=\partial_\mu\mathcal A_\nu-\partial_\nu\mathcal A_\mu-i[\mathcal A_\mu,\mathcal A_\nu],
\end{equation}
which for spatially uniform Rashba coupling is generated purely by the commutator, $F\sim\alpha_R^2$ and $F_{0i}\sim\alpha_R h$. The minimal invariant that is odd in gradients and odd in the field is proportional to $F_{yz}\times\mathcal A\times\mathcal A_0\sim\alpha_R^3h$; combined with the disorder factors of the diffusive gradient expansion controlled by the elastic scattering time $\tau$, this produces the characteristic small parameter $(\alpha_R p_F\tau)^2\cdot(\alpha_R h)$ carried by every Lifshitz invariant obtained below. The non-Abelian gauge structure thus determines the form of all terms derived below.

\subsection{Dyakonov-Perel relaxation gaps of the triplet cooperons}

In a diffusive superconductor the soft modes that build the pairing susceptibility are the cooperons. With spin-orbit coupling the cooperon carries a spin structure: one singlet mode and three triplet modes. The singlet cooperon is protected by particle number and time-reversal-like symmetries and remains gapless (its ``mass'' is $Dq^2-2i\e$), but the triplets precess in the momentum-dependent spin-orbit field and, upon disorder averaging, acquire DP relaxation gaps $\Gamma_t\sim(\alpha_R p_F)^2\tau$. The Lifshitz invariant arises from a second-order process: a singlet fluctuation converts into a triplet through the Zeeman field $h$, propagates as a triplet, and converts back through the field-strength vertex $\sim\alpha_R^3q$. The intermediate triplet propagator is therefore probed at energies $\e\sim T$: if $\Gamma_t\ll T$ the triplet is effectively gapless and the conversion process is limited by temperature, giving Lifshitz couplings $\propto(\alpha_R p_F\tau)^2\alpha_R h/T^2$; if $\Gamma_t\gg T$ the triplet lifetime saturates the process and the couplings become $\propto\alpha_R h\tau/T$, so that the helical wave vector
\begin{equation}
q_0=\frac{4\alpha_R h}{v_F^2}\,\mathfrak g\Big(\frac{\Gamma_t}{4\pi T}\Big),
\qquad \mathfrak g(g\gg1)=1,
\qquad \mathfrak g(g\ll1)=\frac{14\zeta(3)}{\pi^2}g ,
\label{eq:q0-summary}
\end{equation}
no longer depends on the disorder strength in the DP-dominated regime. The full crossover function $\mathfrak g$, Eq.~\eqref{eq:gfun} and Fig.~\ref{fig:crossover}, is obtained in closed form; its two limits correspond to the two families of results previously reported in the literature \cite{DimitrovaFeigelman:2007,HouzetMeyer:2015,IlicBergeret:2022,VBT:2022}. The crossover takes place at $\Gamma_t\sim4\pi T$, i.e., at a spin-orbit splitting $\alpha_Rp_F\sim\sqrt{2\pi T/\tau}$. In the diffusive regime $T\tau\ll1$ this scale is small compared with the disorder broadening $1/\tau$: a Rashba splitting that is weak on the scale of $1/\tau$, in the sense that $\alpha_Rp_F\tau\ll1$, can still exceed the temperature scale and place the system in the relaxation-dominated regime near $T_c$.

\subsection{Nonreciprocal noise: Onsager and the FDT}

Thermodynamic Lifshitz invariants are real and reside in the static free energy. The Keldysh formulation, however, also determines the imaginary (dissipative) part of the pair susceptibility, which defines the kinetic coefficient $\gamma(\bm q;\bm h)$ of the pair field, $\Imy L_R^{-1}(\bm q,\omega)=\omega\,\gamma(\bm q;\bm h)$, and there the same microscopic process produces a term odd in $\bm q$:
\begin{equation}
\gamma(\bm q;\bm h)=\frac{\Imy\, L_R^{-1}(\bm q,\omega)}{\omega}=\frac{\pi}{8T}\big[1+\bm\rho\cdot\bm q\big],
\qquad
\bm\rho=\frac{(\alpha_R p_F\tau)^2}{2T^2}\,[\bm h\times\bm\alpha] ,
\label{eq:ImLR-summary}
\end{equation}
where $\bm\alpha=\alpha_R\hat{\bm x}$ is the Rashba vector along the polar axis (Fig.~\ref{fig:geometry}), i.e., fluctuations with pair momenta parallel and antiparallel to $[\bm h\times\bm\alpha]$ decay at different rates. Such a momentum-odd kinetic coefficient is admissible because $\bm h$ is odd under time reversal: the Onsager relation $\gamma(\bm q;\bm h)=\gamma(-\bm q;-\bm h)$ then permits a term $\propto[\bm h\times\bm\alpha]\cdot\bm q$, while forbidding any $\bm q$-odd term that is even in $\bm h$. The FDT does the rest: in equilibrium the Keldysh component of the pair propagator is fixed by $\Imy L_R^{-1}$, so the Langevin noise power inherits the same directional asymmetry,
\begin{equation}
\big\langle\xi_\Delta(\bm q,\omega)\,\xi^*_\Delta(\bm q,\omega)\big\rangle
=\frac{\pi\omega}{8T\nu}\coth\Big(\frac{\omega}{2T}\Big)\big[1+\bm\rho\cdot\bm q\big].
\label{eq:noise-summary}
\end{equation}
Friction and noise change together. As a consequence the equal-time fluctuation spectrum remains Gibbsian, $\langle|\Delta_{\bm q}|^2\rangle=T/\nu a(\bm q)$, with $a(\bm q)$ the static kernel of the free energy: it is nonreciprocal only through the thermodynamic invariants contained in $a(\bm q)$, it does not depend on $\bm\rho$, and no current flows in equilibrium (Bloch theorem). The kinetic invariant appears in the relaxation dynamics, in the finite-frequency noise spectra, and in the nonlinear response. We call the coupling $\bm\rho$ a kinetic Lifshitz invariant: invisible in thermodynamics, it enters nonlinear fluctuation transport on the same footing as the cubic gradient invariant. In the magnetochiral anisotropy computed in Sec.~\ref{sec:MCA}, for example, the two contribute equally, once both are referred to the helical ground state.

\begin{table}[t]
\centering
\begin{tabular}{llll}
\toprule
Coupling & Meaning & Weak DP ($\Gamma_t\ll4\pi T$) & Strong DP ($\Gamma_t\gg4\pi T$)\\
\midrule
$\bm\Lambda$ & linear Lifshitz invariant &
$\dfrac{7\zeta(3)}{2\pi^2}\dfrac{(\alpha_R p_F\tau)^2}{T^2}\,\bm n$ &
$\dfrac{\pi\tau}{2T}\,\alpha_R h\,\hat{\bm n}$ \\[10pt]
$\Xi$ & cubic gradient invariant &
$\dfrac{\pi D}{16}\dfrac{(\alpha_R p_F\tau)^2\alpha_R h}{T^3}$ &
$\dfrac{7\zeta(3)}{2\pi^2}\dfrac{D\,\alpha_R h\tau}{T^2}$ \\[10pt]
$\bm\rho$ & kinetic (friction/noise) invariant &
$\dfrac{(\alpha_R p_F\tau)^2}{2T^2}\,\bm n$ &
$\dfrac{28\zeta(3)}{\pi^3}\dfrac{\alpha_R h\tau}{T}\,\hat{\bm n}$ \\[10pt]
$\bm{\mathfrak b}$ & quartic (vertex) invariant &
$\dfrac{124\,\zeta(5)}{7\zeta(3)}\dfrac{(\alpha_R p_F\tau)^2}{\pi^2T^2}\,\bm n$ &
(see \ref{app:quartic}) \\[8pt]
\midrule
$q_0$ & helical wave vector & $\dfrac{14\zeta(3)}{\pi^2}\,\dfrac{\Gamma_t}{4\pi T}\dfrac{4\alpha_R h}{v_F^2}$ & $\dfrac{4\alpha_R h}{v_F^2}$ \\
\bottomrule
\end{tabular}
\caption{Summary of the nonreciprocal couplings of the generalized Ginzburg-Landau theory, Eqs.~\eqref{eq:GLfunctional}, \eqref{eq:TDGL}, and \eqref{eq:noisefinal}. Here $\bm n=[\bm h\times\bm\alpha]$ (for $\bm\Lambda$, $\bm\rho$, $\bm{\mathfrak b}$ the magnitude of $\bm n$, $|\bm n|=\alpha_R h$, is included in the quoted vectors), $\Gamma_t=2(\alpha_R p_F)^2\tau$ is the DP rate, and $D=v_F^2\tau/2$. The entries $\bm\Lambda$, $\Xi$, $\bm\rho$, and $q_0$ follow from the single kernel $\mathcal K(x;\Gamma_t)$ of Eq.~\eqref{eq:Kkernel} and its derivatives, with the two columns obtained from their limits. The quartic-vertex invariant $\bm{\mathfrak b}$ requires the full three-momentum structure of the vertex and is fixed by the diagrammatic evaluation of Ref.~\cite{Hasan:2025} (see Sec.~\ref{sec:quartic} and \ref{app:quartic}). In the strong-DP column the three entries $\bm\Lambda$, $\Xi$, $\bm\rho$ are not independent: $\Xi=2\Lambda\alpha_4/\xi^2_{GL}$ and $\rho=-\Lambda \rho_2/\xi^2_{GL}$ [Eqs.~\eqref{eq:evenquartic} and \eqref{eq:w2}], since all three follow from a single shift of the pair momentum (Sec.~\ref{sec:master}).}
\label{tab:couplings}
\end{table}

%==============================================================================================================================================================================================
%==============================================================================================================================================================================================
%==============================================================================================================================================================================================
%==============================================================================================================================================================================================
%==============================================================================================================================================================================================
\section{Keldysh sigma model for the disordered Rashba superconductor}
\label{sec:formalism}

\subsection{Model and gauge structure}
\label{sec:model}

We consider a two-dimensional electron gas with Rashba spin-orbit coupling, an in-plane Zeeman field, an attractive interaction in the $s$-wave singlet channel, and potential disorder:
\begin{equation}
H=\frac{p^2}{2m}+\alpha_R\,[\bm p\times\hat{\bm x}]\cdot\bm s
+\bm h\cdot\bm s+U_{\rm dis}(\bm r)-\lambda\,\psi^\dagger_\uparrow\psi^\dagger_\downarrow\psi_\downarrow\psi_\uparrow .
\label{eq:H}
\end{equation}
In our convention the electrons move in the $(y,z)$ plane and $\hat{\bm x}$ is the polar (Rashba) axis normal to it; $\bm s$ are the spin Pauli matrices; and the Zeeman field $\bm h=h\hat{\bm z}$ lies in the plane. The disorder potential is Gaussian and short-ranged, with elastic scattering time $\tau$; we work in the diffusive regime $T\tau\ll1$ and treat the spin-orbit splitting as small on the scale of the disorder broadening, $\alpha_R p_F\tau\ll1$. No restriction is placed on the ratio $\Gamma_t/T$, where
\begin{equation}
\Gamma_t=2(\alpha_R p_F)^2\tau
\label{eq:Gammat}
\end{equation}
is the DP spin-relaxation rate; this ratio will emerge as the parameter controlling the disorder dependence of the magnetoelectric couplings of the quadratic sector. Note that the window $\alpha_R p_F\tau\ll1$ includes both $\Gamma_t\ll T$ and $\Gamma_t\gg T$, since $\Gamma_t/T=2(\alpha_R p_F\tau)^2/(T\tau)$ and $T\tau\ll1$.

As anticipated in Sec.~\ref{sec:overview}, we encode the spin structure of Eq.~\eqref{eq:H} as a background SU(2) gauge field \cite{BergeretTokatly:2014,VBT:2021}. Writing $\alpha_R[\bm p\times\hat{\bm x}]\cdot\bm s=\alpha_R(p_zs_2-p_ys_3)$, the kinetic and spin terms combine into $(\bm p+\bm{\mathcal A})^2/2m+\mathcal A_0$ with
\begin{equation}
\mathcal A_0=h\hat s_3,\qquad
\mathcal A_y=-m\alpha_R\,\hat s_3,\qquad
\mathcal A_z=m\alpha_R\,\hat s_2 ,
\label{eq:su2}
\end{equation}
up to an irrelevant constant. The associated field strength $F_{\mu\nu}=\partial_\mu\mathcal A_\nu-\partial_\nu\mathcal A_\mu-i[\mathcal A_\mu,\mathcal A_\nu]$ has, in our geometry, only two nonvanishing components, both generated by the commutator:
\begin{equation}
F_{yz}=2m^2\alpha_R^2\,\hat s_1,
\qquad
F_{0z}=-2m\alpha_R h\,\hat s_1 .
\label{eq:Fcomp}
\end{equation}
The spatial component $F_{yz}$ is the ``spin magnetic field'' responsible for the intrinsic spin-Hall physics of the Rashba model; the temporal component $F_{0z}$ is a ``spin electric field'' and will not be needed at the order we work to (it would contribute magnetoelectric terms at higher order in $\omega$ and $h$). The gauge-covariance argument of Sec.~\ref{sec:overview} tells us to expect all Lifshitz invariants to be assembled from $F_{yz}$, one power of $\mathcal A_{y,z}$ (to supply the third power of $\alpha_R$ together with a spatial gradient), and one power of $\mathcal A_0$ (to break time reversal), i.e., to scale as $\alpha_R^3h\times(\text{disorder factors})$.

\subsection{The sigma-model action}
\label{sec:action}

The low-energy, long-wavelength physics of a disordered superconductor is captured by the Keldysh nonlinear sigma model \cite{FLS:2000,KamenevAndreev:1999,KamenevBook}. Its degree of freedom is a matrix field $Q_{tt'}(\bm r)$, obeying the nonlinear constraint $Q^2=1$, which acts in the product of Keldysh ($\hat\sigma_i$), Nambu ($\hat\tau_i$), and spin ($\hat s_i$) spaces and carries two time (or energy) arguments. Physically, $Q$ is the disorder-averaged, spatially smoothed electron Green function at coincident points; its soft fluctuations around the metallic saddle point are the diffusons and cooperons of the diagrammatic technique. The superconducting order parameter $\Delta(\bm r,t)$ enters as a Hubbard--Stratonovich field decoupling the BCS interaction, and has two components on the Keldysh contour $\Delta^{c}$ and $\Delta^{q}$ (classical and quantum), defined as the half-sum and the half-difference of the fields on the forward and backward branches, $\Delta^{c,q}=(\Delta_+\pm\Delta_-)/2$, the doubling being the hallmark of the closed-time-path formulation. The classical component is the physical order parameter, while the quantum component generates equations of motion, dissipation, and (through the quadratic term) noise \cite{Kamenev:2009}.

For our model the action, including the leading field-strength term obtained at fourth order of the gradient expansion of Refs.~\cite{VBT:2021,VBT:2022}, reads
\begin{equation}
S[Q,\Delta]=\frac{i\pi\nu}{4}\Tr\Big[
\frac{4i}{\pi\lambda}\Delta\hat\sigma_x\Delta
+D(\tilde\nabla_kQ)^2
+4i\big(i\hat\tau_3\partial_t+\mathcal A_0\hat\tau_3+\Delta\big)Q
+\varkappa_g\,F_{ij}\,Q\tilde\nabla_iQ\tilde\nabla_jQ\Big],
\label{eq:S}
\end{equation}
where $\varkappa_g=Dl/p_F$, $\tilde\nabla_k=\nabla_k-i[\mathcal A_k,\,\cdot\,]$ is the SU(2)-covariant derivative, $D=v_F^2\tau/2$ is the diffusion constant, $l=v_F\tau$, $\nu$ is the density of states at the Fermi level, and $\Tr$ includes the full matrix trace together with the time/energy and spatial integrations. Electromagnetic potentials, not written explicitly, enter through the same covariant derivative in the charge channel and are restored at the end in the standard way \cite{LevchenkoKamenev:2007}.

Let us comment on the physical content of each term. The first term is the decoupled BCS interaction; $\hat\sigma_x$ in Keldysh space pairs classical with quantum components, as required by the causality structure of the Keldysh technique (no $\Delta^{c}\Delta^{c}$ term may appear). The second term is the diffusive stiffness: for the spin-singlet sector it describes ordinary charge diffusion, while for the spin-triplet sector the covariant derivative embodies the precession of the electron spin in the Rashba field and, at second order, the DP relaxation. The third term contains the dynamics ($\partial_t$), the Zeeman coupling, and the linear source for the pairing field. The last term is the leading non-Abelian field-strength contribution \cite{VBT:2021,VBT:2022}: it is the sigma-model image of the intrinsic spin-Hall coupling, and it is the only place, at this order, where the gauge-invariant $F_{ij}$ enters. 
\footnote{At this order the gradient expansion of Ref.~\cite{VBT:2022} contains one additional term, denoted $S'_{4,1}$ there, which is nominally of the same order as the $\varkappa_g$ term but does not involve the field strength $F_{ij}$. Consequently, it does not contribute to the TDGL theory of a spin-singlet superconductor and can be omitted here.}
Since every Lifshitz invariant computed below is directly proportional to $\varkappa_g$, the numerical prefactor of this term should be kept in mind when comparing with other conventions.

\subsection{Cooperon parametrization}
\label{sec:cooperons}

In the normal metal, at $\Delta=\mathcal A_\mu=0$, the saddle point of the action \eqref{eq:S} is
\begin{equation}
\Lambda_\e=\begin{pmatrix}
1 & 2f_\e\\ 0 & -1
\end{pmatrix}_K\hat\tau_3 ,
\qquad f_\e=\tanh\frac{\e}{2T},
\label{eq:Lambda}
\end{equation}
whose retarded-advanced-Keldysh triangular structure encodes causality, with the equilibrium distribution $f_\e$ in the off-diagonal (Keldysh) block. Fluctuations on the $Q^2=1$ manifold are parametrized by rotations
\begin{equation}
Q=U\,e^{-w/2}\,\hat\sigma_3\hat\tau_3\,e^{w/2}\,U^{-1}
=U\,\hat\sigma_3\hat\tau_3\,e^{w}\,U^{-1},
\qquad
U=U^{-1}=\begin{pmatrix}1 & f_\e\\ 0 & -1\end{pmatrix}_K ,
\label{eq:param}
\end{equation}
where $w$ anticommutes with $\hat\sigma_3\hat\tau_3$ (the second equality follows from this anticommutation and makes the expansion in $w$ transparent: $Q=U\hat\sigma_3\hat\tau_3(1+w+w^2/2+\dots)U^{-1}$). The generators that couple linearly to the pairing field are the cooperons, residing in the Nambu off-diagonal blocks. Suppressing the diffuson generators (which couple to $\Delta$ only at cubic order and play no role in what follows), we write
\begin{equation}
w=w_\mu\hat s_\mu,
\qquad
w_\mu=\begin{pmatrix}
c_\mu\hat\tau_+-c^*_\mu\hat\tau_- & 0\\
0 & \bar c_\mu\hat\tau_+-\bar c^{\,*}_\mu\hat\tau_-
\end{pmatrix}_K ,
\label{eq:w}
\end{equation}
with $\hat\tau_\pm=(\hat\tau_1\pm i\hat\tau_2)/2$. The index $\mu=0,1,2,3$ corresponds to the spin structure of the Cooper pair: $\mu=0$ is the singlet channel and $\mu=1,2,3$ are the triplets. The two Keldysh blocks are physically distinct: expanding the time-derivative term of the action one finds that $c_\mu(\bm q;\e,\e')$ propagates with the retarded cooperon kernel $\propto Dq^2-i(\e+\e')$ while $\bar c_\mu$ propagates with the advanced one $\propto Dq^2+i(\e+\e')$. This retarded/advanced doubling is what will eventually produce the retarded, advanced, and Keldysh components of the pair fluctuation propagator.

\subsection{Gaussian action of the cooperons}
\label{sec:gaussianaction}

We now expand the action \eqref{eq:S} to second order in the cooperon generators. This step involves lengthy but elementary traces in the $8\times8$ Keldysh$\otimes$Nambu$\otimes$spin space, see (\ref{app:gaussian}). Arranging the quadratic form as
\begin{equation}
S_c= i\pi\nu\int\!\frac{d\e\,d\e'}{4\pi^2}\int\!\frac{d^2q}{(2\pi)^2}\;
c_\mu(\bm q;\e,\e')\,\mathcal C_{\mu\nu}(\bm q,\e+\e')\,c^*_\nu(-\bm q;\e',\e)
+(\bar c\ \text{sector})+(\text{sources}),
\end{equation}
we find, with $a\equiv Dq^2-i(\e+\e')$, 
\begin{equation}
\mathcal C(\bm q,\e+\e')=
\begin{pmatrix}
a & 0 & -4i\kappa_0q_z & -2ih+4i\kappa_0q_y\\[2pt]
0 & a+2\Gamma_t & 4iDm\alpha_R\,q_y & 4iDm\alpha_R\,q_z\\[2pt]
-4i\kappa_0q_z & 4iDm\alpha_R\,q_y & a+\Gamma_t & 0\\[2pt]
-2ih+4i\kappa_0q_y & 4iDm\alpha_R\,q_z & 0 & a+\Gamma_t
\end{pmatrix}_{N\otimes S},
\label{eq:Amatrix}
\end{equation}
where the field-strength coupling constant is
\begin{equation}
\kappa_0\equiv m^3\alpha_R^3\varkappa_g=\frac{\alpha_R}{2}\,(\alpha_R p_F\tau)^2 .
\label{eq:kappa0}
\end{equation}
The advanced sector is obtained by $a\to\bar a=Dq^2+i(\e+\e')$ with all magnetoelectric couplings ($h$ and $\kappa_0$ entries) reversed in sign. Equation \eqref{eq:Amatrix} is the starting point of technical derivation, so we discuss its entries one by one.

(i) \emph{Diffusion pole (entry $00$).} The singlet cooperon is gapless: its propagator $1/a$ has the diffusion pole that, upon attaching the BCS coupling, produces the logarithmic Cooper instability. The gaplessness is protected: the singlet is a scalar under spin rotations, so the covariant derivative acts on it as an ordinary derivative, $[\mathcal A_k,\hat s_0]=0$. This is the sigma-model expression of the Anderson theorem.

(ii) \emph{Dyakonov-Perel gaps (diagonal, triplet block).} Expanding $D(\tilde\nabla Q)^2$ produces, besides $Dq^2$, the commutator-squared term $-D[\mathcal A_k,Q][\mathcal A_k,Q]$, which gives the triplet cooperons finite relaxation rates even at $\bm q=0$: $\Gamma_2=\Gamma_3=\Gamma_t=4Dm^2\alpha_R^2$ for the two in-plane triplets and $\Gamma_1=2\Gamma_t$ for the triplet polarized along the Rashba axis. This factor-of-two anisotropy is the standard DP pattern: the $\hat s_1$ triplet is relaxed by both in-plane components of the Rashba field, while $\hat s_2$ and $\hat s_3$ are each relaxed by only one. Since $D=v_F^2\tau/2$, we have $\Gamma_t=2(\alpha_R p_F)^2\tau$: the relaxation rate is proportional to the scattering time and therefore decreases with increasing disorder, the motional-narrowing behavior characteristic of the DP mechanism.

(iii) \emph{Triplet precession (entries $12$, $13$).} The cross term $-2iD\,\nabla Q\,[\mathcal A,Q]$ describes the coherent precession of a moving triplet in the Rashba field and couples $c_1$ to $c_{2,3}$ with matrix elements $\propto Dm\alpha_R q$. It does not couple to the singlet --- at linear order in $\mathcal A$ the singlet is gauge-protected, in accord with Sec.~\ref{sec:model}. These entries feed into the singlet block of the inverse only at relative order $(Dm\alpha_R q)^2/\Gamma_t^2\sim Dq^2/\Gamma_t$ and affect none of the leading Lifshitz couplings. 

(iv) \emph{Zeeman coupling (entries $03$, $30$).} The term $4ih\Tr[\hat\tau_3\hat s_3Q]$, expanded to second order in $w$, couples the singlet to the triplet polarized along $\bm h$ with the symmetric matrix element $-2ih$. 
The symmetry (as opposed to antisymmetry) of this coupling is fixed by the physics of Zeeman pair breaking, as follows.
Neglecting spin-orbit terms, the singlet block of the inverse kernel is
\begin{equation}
\mathcal C^{-1}_{00}=\Big[a+\frac{4h^2}{a+\Gamma_t}\Big]^{-1}
\;\xrightarrow[\Gamma_t\to0]{}\;
\frac{a}{a^2+4h^2}
=\frac12\Big[\frac{1}{Dq^2-2i(\tilde\e+h)}+\frac{1}{Dq^2-2i(\tilde\e-h)}\Big],
\label{eq:zeemanpoles}
\end{equation}
with $\tilde\e=(\e+\e')/2$, i.e., the pair susceptibility of electrons whose energies are Zeeman-shifted by real $\pm h$. This is the familiar singlet pair-breaking structure, which integrates to the standard dirty-limit result [Eq.~\eqref{eq:hsquare} below]. An antisymmetric coupling would produce $a/(a^2-4h^2)$, with unphysical poles at imaginary Zeeman shifts. At finite $\Gamma_t$, Eq.~\eqref{eq:zeemanpoles} interpolates to $[a+4h^2\tau_{t}]^{-1}$ with $\tau_t=\Gamma_t^{-1}$: fast DP relaxation averages the pair-breaking away, the well-known spin-orbit protection of paramagnetically limited superconductors \cite{KLB:1975}.

(v) \emph{Field-strength coupling (entries $02$, $03$).} Finally, the last term of \eqref{eq:S} contributes at Gaussian order through $Q\to\Lambda$ with one covariant-derivative commutator and one plain gradient, producing singlet-triplet matrix elements linear in momentum,
\begin{equation}
\mathcal C^{(F)}_{03}=\mathcal C^{(F)}_{30}=+4i\kappa_0q_y,
\qquad
\mathcal C^{(F)}_{02}=\mathcal C^{(F)}_{20}=-4i\kappa_0q_z .
\end{equation}
These are again symmetric in the spin indices and thus odd under $\bm q\to-\bm q$, as a Lifshitz coupling must be. The combination $\kappa_0=m^3\alpha_R^3\varkappa_g$ was already anticipated by dimensional analysis in Sec.~\ref{sec:overview}: two powers of $\alpha_R$ from $F_{yz}$ and one from the commutator $[\mathcal A_i,\,\cdot\,]$, and the disorder factor $\varkappa_g=Dl/p_F$ yielding $\kappa_0=\tfrac{\alpha_R}{2}(\alpha_R p_F\tau)^2$.

One more structural remark will be important for the dynamics. Between the retarded ($c$) and advanced ($\bar c$) sectors both the Zeeman and the field-strength magnetoelectric couplings reverse sign together. Their product, which generates the Lifshitz invariants, is therefore identical in the two sectors. This guarantees the relation $L_A^{-1}(\bm q,\omega)=[L_R^{-1}(\bm q,\omega)]^*$ after including all nonreciprocal terms, a consistency requirement of the Keldysh structure.

Because the singlet talks to the triplet sector only through the $h$ and $\kappa_0q$ entries, the singlet block of $\mathcal C^{-1}$ is obtained by inverting the $\{0,2,3\}$ block:
\begin{equation}
\mathcal C^{-1}_{00}(\bm q,\tilde\e)
=\bigg[a+\frac{(2h-4\kappa_0q_y)^2+16\kappa_0^2q_z^2}{a+\Gamma_t}\bigg]^{-1}
\simeq
\frac1a
-\frac{4h^2}{a^2(a+\Gamma_t)}
+\frac{16h\kappa_0q_y}{a^2(a+\Gamma_t)}+\ldots.
\label{eq:Ainv}
\end{equation}
where the higher-order terms in the expansion are $O(h^4,\,h^3\kappa_0q,\,\kappa_0^2q^2)$. The cross term $\propto h\kappa_0q_y$ arises from conversion through one Zeeman vertex and one field-strength vertex, with a triplet propagating in between, and produces the Lifshitz invariants. It requires broken parity ($\kappa_0$) and broken time reversal ($h$) simultaneously, and it is odd in $\bm q$, as it must be. Note the triplet denominator $a+\Gamma_t$: the DP gap regulates the intermediate triplet propagation and is the microscopic origin of the crossover discussed below.

\subsection{Order-parameter sources and Gaussian integration}
\label{sec:sources}

To first order in $w$ the source term of the action reads
\begin{equation}
\frac{i\pi\nu}{4}\,\Tr\big[4i\Delta Q\big]^{(1)}
\;\propto\;
\big(\Delta^{c}_\mu+f_\e\,\Delta^{q}_\mu\big)\,c^*_\mu
+\big(\Delta^{c}_\mu-f_{\e'}\,\Delta^{q}_\mu\big)\,\bar c^{\,*}_\mu
-\text{c.c.},
\label{eq:sources}
\end{equation}
with the characteristic distribution-function dressing of the quantum component: the retarded sector couples to $\Delta^{c}+f_\e\Delta^q$ (distribution at the first energy argument), the advanced sector to $\Delta^{c}-f_{\e'}\Delta^q$. This asymmetry is what converts, after Gaussian integration, the retarded/advanced cooperon kernels into the retarded, advanced, and Keldysh blocks of the pair propagator with the correct equilibrium relations among them.

Integrating out the cooperons is now a Gaussian exercise. The quadratic action of the order parameter takes the form
\begin{equation}
S_{GL}=2\nu\!\int\!\frac{d\omega}{2\pi}\!\int\!\frac{d^2q}{(2\pi)^2}
\Big[\Delta^{*q}\,L_R^{-1}\,\Delta^{c}
+\Delta^{*c}\,L_A^{-1}\,\Delta^{q}
+\Delta^{*q}\,L_K^{-1}\,\Delta^{q}\Big],
\label{eq:SGL}
\end{equation}
with
\begin{align}
L_{R}^{-1}(\bm q,\omega)&=-\frac1\lambda
-i\!\int\! d\e\;
f_{\e-\omega/2}\;\mathcal C^{-1}_{00}(\bm q,\e),
\qquad
L_A^{-1}=\big[L_R^{-1}\big]^*,
\label{eq:Ldefs}\\
L_K^{-1}(\bm q,\omega)&=-i\!\int\! d\e\;
f_{\e-\omega/2}\,f_{\e+\omega/2}\;\mathcal C^{-1}_{00}(\bm q,\e) .
\label{eq:LKdef}
\end{align}
The overall factor in \eqref{eq:SGL} is fixed by the requirement that in the static limit the action reduce to the difference of the Ginzburg-Landau free energies on the two branches of the contour, $S_{GL}|_{\omega\to0}=-\int dt\,\{F[\Delta_+]-F[\Delta_-]\}$, with $F$ the free energy of Sec.~\ref{sec:freeenergy} and $\nu$ the density of states per spin; with the half-sum convention for $\Delta^{c,q}$ this produces the factor $2\nu$ in front of $\Delta^{*q}L_R^{-1}\Delta^c+{\rm c.c.}$ with $L_R^{-1}(\bm q,0)=-a(\bm q)$. All normalizations below, in particular the equal-time fluctuation amplitude \eqref{eq:gibbs} and the noise correlator \eqref{eq:noisefinal}, refer to this convention. At $\alpha_R=h=0$ the action coincides with that of Ref.~\cite{LevchenkoKamenev:2007} [Eqs.~(44)--(46) there], where the kinetic coefficient $\pi/8T$ is factored out of the kernel. The absence of a $\Delta^{*c}\Delta^{c}$ term verifies the Keldysh normalization. In equilibrium the three blocks are not independent. Using the hyperbolic identity
\begin{equation}
f_{\e-\omega/2}\,f_{\e+\omega/2}-1
=\coth\Big(\frac{\omega}{2T}\Big)\big[f_{\e-\omega/2}-f_{\e+\omega/2}\big],
\label{eq:tanhidentity}
\end{equation}
the Keldysh block becomes $L_K^{-1}=\coth(\omega/2T)[L_R^{-1}-L_A^{-1}]-i\int d\e\,\mathcal C^{-1}_{00}(\bm q,\e)$. The last term requires a comment. For a symmetric energy cutoff the integral of the retarded function $\mathcal C^{-1}_{00}$ does not vanish (for the bare diffusion pole it equals $\pi/2$), and it would formally add a frequency-independent constant to the noise kernel. The term originates from the constant $(-1)$ in \eqref{eq:tanhidentity}, i.e., from the equal-time, $\delta(t-t')$, part of the kernel, and in the Keldysh technique the energy integral of a retarded function at coincident times is evaluated with the causal regularization $\int d\e\,e^{i\e0^+}\mathcal C^{-1}_{00}(\bm q,\e)=0$, which closes the contour in the half-plane where $\mathcal C^{-1}_{00}$ is analytic. This is the convention $\theta(0)=0$ for retarded functions that guarantees the normalization $Z=1$ of the Keldysh partition function \cite{KamenevBook,LevchenkoKamenev:2007}. With it we obtain the relation
\begin{equation}
L_K^{-1}(\bm q,\omega)
=\coth\Big(\frac{\omega}{2T}\Big)\Big[L_R^{-1}(\bm q,\omega)-L_A^{-1}(\bm q,\omega)\Big]
=2i\coth\Big(\frac{\omega}{2T}\Big)\,\Imy\,L_R^{-1}(\bm q,\omega).
\label{eq:FDT}
\end{equation}
This is the fluctuation-dissipation theorem at the level of the pair susceptibility. Equation \eqref{eq:FDT} says that in equilibrium the Keldysh (noise) block contains no information beyond $\Imy L_R^{-1}$: every term of the latter, including the momentum-odd one found below, enters the noise with the universal weight $\coth(\omega/2T)$. The content of $\Imy L_R^{-1}$ itself is a matter of microscopic calculation, and in a noncentrosymmetric system it turns out to be nonreciprocal. Out of equilibrium, when $f_\e$ is not the thermal function, the identity \eqref{eq:tanhidentity} no longer collapses the Keldysh block, and independent nonreciprocal noise appears; we return to this point in Sec.~\ref{sec:conclusions}.

%==============================================================================================================================================================================================
%==============================================================================================================================================================================================
%==============================================================================================================================================================================================
%==============================================================================================================================================================================================
%==============================================================================================================================================================================================
\section{The generalized Ginzburg--Landau functional}
\label{sec:GL}

\subsection{Energy integrals and the master kernel}
\label{sec:integralsmain}

The energy integrals in Eq.~\eqref{eq:Ldefs} are performed in closed form. Shifting $\e\to\e+\omega/2$ moves the frequency into the diffusion pole, $a\to b-2i\e$ with $b=Dq^2-i\omega$, and every required integral reduces to
\begin{equation}
J_n(b)=\int d\e\,\tanh\Big(\frac{\e}{2T}\Big)\frac{1}{(b-2i\e)^n} .
\end{equation}
Expanding $\tanh x=\sum_{k\ge0}2x/(x^2+x_k^2)$ over its Matsubara poles $x_k=\pi(k+\tfrac12)$ and closing the contour in the half-plane free of the (retarded) diffusion pole, one finds for $n\ge2$ (\ref{app:integrals})
\begin{equation}
J_n=\frac{2T}{(4T)^n}\,\frac{2\pi i\,(-1)^n}{\pi^n (n-1)!}\,
\psi^{(n-1)}\Big(\frac12+x\Big),
\qquad
x\equiv\frac{b}{4\pi T}=\frac{Dq^2-i\omega}{4\pi T},
\label{eq:Jn}
\end{equation}
with $\psi^{(n)}$ the polygamma functions, while the logarithmically divergent $J_1$ is regularized by the BCS cutoff into the standard combination $\ln(T_c/T)-\psi(\tfrac12+x)+\psi(\tfrac12)$. For the magnetoelectric terms of Eq.~\eqref{eq:Ainv} the additional triplet denominator appears, and the required master integral is
\begin{align}
\mathcal K(x;\Gamma_t)\equiv&
-i\!\int\! d\e\,\frac{\tanh(\e/2T)}{(b-2i\e)^2\,(b+\Gamma_t-2i\e)}\nonumber \\ 
&=\frac{1}{4\pi T\,\Gamma_t}
\bigg[\psi'\Big(\frac12+x\Big)
-\frac{\psi\big(\frac12+x+g\big)-\psi\big(\frac12+x\big)}{g}\bigg],
\label{eq:Kkernel}
\end{align}
where $g=\Gamma_t/4\pi T$, obtained by partial fractions, $[a^2(a+\Gamma)]^{-1}=(\Gamma a^2)^{-1}-(\Gamma^2a)^{-1}+[\Gamma^2(a+\Gamma)]^{-1}$, and Eq.~\eqref{eq:Jn}. Its limits are
\begin{equation}
\mathcal K(x;\Gamma_t)\to
\begin{cases}
\displaystyle
-\frac{\psi^{(2)}\big(\frac12+x\big)}{32\pi^2T^2}
=\frac{7\zeta(3)}{16\pi^2T^2}+O(x),
& \Gamma_t\ll4\pi T,
\\[14pt]
\displaystyle
\frac{\psi'\big(\frac12+x\big)}{4\pi T\,\Gamma_t}
=\frac{\pi}{8T\,\Gamma_t}+O(x),
& \Gamma_t\gg4\pi T .
\end{cases}
\label{eq:Klimits}
\end{equation}
Physically, for weak DP relaxation the intermediate triplet in the singlet to triplet to singlet conversion process lives long compared with the thermal coherence time $1/T$, and the process is cut off thermally ($\mathcal K\sim1/T^2$); for strong relaxation, on the other hand, the triplet lifetime $\Gamma_t^{-1}$ provides the cutoff ($\mathcal K\sim 1/T\Gamma_t$).

\subsection{Retarded pair propagator}
\label{sec:master}

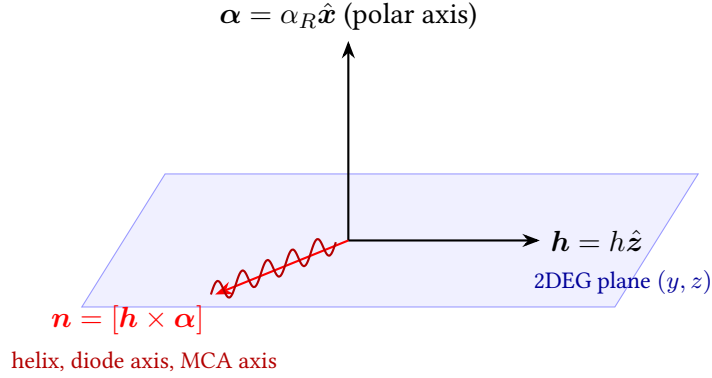
\begin{figure}[t]
\centering
\begin{tikzpicture}[scale=1.1,>=Stealth]
% plane
\draw[fill=blue!6,draw=blue!40] (-3.2,-1.4) -- (3.2,-1.4) -- (4.2,0.2) -- (-2.2,0.2) -- cycle;
\node[blue!60!black] at (3.3,-1.1) {\footnotesize 2DEG plane $(y,z)$};
% axes
\draw[->,thick] (0,-0.6) -- (0,1.8) node[above] {$\bm\alpha=\alpha_R\hat{\bm x}$ (polar axis)};
\draw[->,thick] (0,-0.6) -- (2.3,-0.6) node[right] {$\bm h=h\hat{\bm z}$};
\draw[->,thick,red] (0,-0.6) -- (-1.6,-1.25) node[below left] {$\bm n=[\bm h\times\bm\alpha]$};
% helix
\draw[domain=-1.5:0,smooth,samples=60,variable=\t,red!70!black,thick]
 plot ({\t-0.15},{-0.62+0.42*\t+0.13*sin(1200*\t)});
\node[red!70!black] at (-2.45,-2.05) {\footnotesize helix, diode axis, MCA axis};
\end{tikzpicture}
\caption{Geometry of the minimal model. The electrons move in the $(y,z)$ plane normal to the Rashba (polar) axis $\bm\alpha$ in an in-plane Zeeman field $\bm h$. All nonreciprocal couplings, such as the helical modulation $\bm q_0$, the diode axis, and the drift vector $\bm\rho$ of the kinetic Lifshitz invariant, point along the in-plane vector $\bm n=[\bm h\times\bm\alpha]$, the only polar in-plane vector odd in $\bm h$ allowed by the $C_{\infty v}$ symmetry.}
\label{fig:geometry}
\end{figure}

Combining Eqs.~\eqref{eq:Ainv}, \eqref{eq:Ldefs}, \eqref{eq:Jn}, and \eqref{eq:Kkernel}, we find
\begin{equation}
L_R^{-1}(\bm q,\omega)=
\ln\frac{T_c}{T}
-\psi\Big(\frac12+x\Big)+\psi\Big(\frac12\Big)
-\Big[4h^2-16\,h\kappa_0q_y\Big]\,\mathcal K(x;\Gamma_t),
\label{eq:LRmaster}
\end{equation}
Expanding the polygamma functions, Eq.~\eqref{eq:LRmaster} generates the magnetoelectric couplings of the quadratic sector with their full disorder dependence.
The resulting expressions can be put in coordinate-free form by noting that $h\kappa_0q_y=\tfrac12(\alpha_R p_F\tau)^2\,\bm h\cdot[\bm\alpha\times\bm q]$ with $\bm\alpha=\alpha_R\hat{\bm x}$, such that the Lifshitz invariant is \(\propto\bm n\cdot\bm q\) with a preferred direction determined by
\begin{equation}
\bm n\equiv[\bm h\times\bm\alpha] ,
\end{equation}
in-plane and perpendicular to the Zeeman field (see Fig.~\ref{fig:geometry}) as dictated by the $C_{\infty v}$ symmetry of the Rashba model, since $\bm n$ is the only polar in-plane vector that is odd in $\bm h$. We now consider some limiting expressions.

(i) \emph{Weak-DP regime}, $\Gamma_t\ll4\pi T$. Expanding \eqref{eq:LRmaster} in $x$ and using $\psi'(\tfrac12)=\pi^2/2$, $\psi^{(2)}(\tfrac12)=-14\zeta(3)$, $\psi^{(3)}(\tfrac12)=\pi^4$, we have
\begin{equation}
L_R^{-1}=
\frac{T_c-T}{T}
+\frac{i\pi\omega}{8T}\big[1+\bm\rho\cdot\bm q\big]
-\frac{\pi D}{8T}\,q^2
-\frac{7\zeta(3)\,h^2}{4\pi^2T^2}
+\bm\Lambda\cdot\bm q
-\Xi\,(\hat{\bm n}\cdot\bm q)\,q^2
+\dots ,
\label{eq:LRweak}
\end{equation}
with
\begin{equation}
\bm\Lambda=\frac{7\zeta(3)\,(\alpha_R p_F\tau)^2}{2\pi^2T^2}\,\bm n,
\qquad
\Xi=\frac{\pi D\,(\alpha_R p_F\tau)^2\,\alpha_R h}{16\,T^3},
\qquad
\bm\rho=\frac{(\alpha_R p_F\tau)^2}{2T^2}\,\bm n .
\label{eq:couplingsweak}
\end{equation}
The first three terms of \eqref{eq:LRweak} are the standard dirty-limit GL expansion with $\xi_{GL}^2=\pi D/8T$ and $\tau_{GL}^{-1}=8(T-T_c)/\pi$. The fourth is the paramagnetic suppression. $\bm\Lambda$ is the linear Lifshitz invariant, $\Xi$ is the cubic one, and the dynamical term $\propto\bm\rho\cdot\bm q$, which has no analog in a free-energy treatment, is the kinetic Lifshitz invariant discussed in Sec.~\ref{sec:TDGL}. Note that all three odd couplings carry factors of $(\alpha_R p_F\tau)^2$: in this regime disorder suppresses the magnetoelectric response, and the suppression is quadratic because, as explained in Sec.~\ref{sec:overview}, the invariants are built from the field strength rather than from the potential.

(ii) \emph{Strong-DP regime}, $\Gamma_t\gg4\pi T$. Here, using $\kappa_0/\Gamma_t=\alpha_R\tau/4$, we have
\begin{equation}
L_R^{-1}\Big|_{\rm Lifshitz}
=\frac{16h\kappa_0q_y}{4\pi T\,\Gamma_t}\,\psi'\Big(\frac12+x\Big)
=\frac{\pi\tau}{2T}\,\bm h\cdot[\bm\alpha\times\bm q]\,
\frac{\psi'\big(\frac12+x\big)}{\psi'\big(\frac12\big)} ,
\label{eq:LRstrong}
\end{equation}
linear in $\tau$ and first order in $\alpha_R$ (times the geometric factor $\alpha_R h$ hidden in the mixed product); simultaneously the paramagnetic term saturates at $-\pi h^2/2T\Gamma_t$, the spin-orbit-protected pair breaking \cite{KLB:1975}. The ratio of the Lifshitz term to the gradient term is now disorder independent, which is the origin of the universal (i.e., disorder-independent) Cooper pair momentum in the helical superconducting state discussed next. Expanding \eqref{eq:LRstrong} to first order in $x$ with $\psi'(\tfrac12+x)/\psi'(\tfrac12)=1-(28\zeta(3)/\pi^2)\,x+O(x^2)$ yields the strong-DP cubic invariant quoted in Table~\ref{tab:couplings}, $\Xi=7\zeta(3)D\alpha_R h\tau/2\pi^2T^2$. A remark on the sign: since $\partial_x\mathcal K(x;\Gamma_t)<0$ for all $g$ [$\mathcal K$ is built from $-\psi^{(2)}(\tfrac12+x)$ at small $g$ and from $\psi'(\tfrac12+x)$ at large $g$, both decreasing functions of $x$], the cubic invariant $\Xi=-16h\kappa_0(D/4\pi T)\,\partial_x\mathcal K|_{x=0}$ is positive at all $g$, with the same orientation as $\bm\Lambda$ and $\bm\rho$; disorder changes its magnitude but not its sign.

The strong-DP limit has one further property, which turns out to control the nonreciprocal observables of Sec.~\ref{sec:apps}. Writing the reciprocal part of \eqref{eq:LRmaster} as $A(x)=\ln(T_c/T)-\psi(\tfrac12+x)+\psi(\tfrac12)$, the limit $\mathcal K\to\psi'(\tfrac12+x)/4\pi T\Gamma_t$ makes the odd term proportional to the derivative of the even one, $16h\kappa_0q_y\,\mathcal K(x)\to-(16h\kappa_0q_y/4\pi T\Gamma_t)\,\partial_xA(x)$, so that to first order in $h$, and at all $\bm q$ and $\omega$,
\begin{equation}
L_R^{-1}(\bm q,\omega)=A\Big(\frac{D(\bm q-q_0\hat{\bm n})^2-i\omega}{4\pi T}\Big)+O(h^2),
\qquad
q_0=\frac{8h\kappa_0}{D\Gamma_t}=\frac{4\alpha_R h}{v_F^2}:
\label{eq:removability}
\end{equation}
the complete retarded kernel, statics and dynamics alike, is the $h=0$ kernel evaluated at the shifted pair momentum. Consequently the strong-DP entries of Table~\ref{tab:couplings} are not independent: expanding \eqref{eq:removability} one finds $\Xi=4\alpha_4q_0=2\Lambda\alpha_4/\xi^2_{GL}$ and $\rho=-2\rho_2q_0=-\Lambda\rho_2/\xi^2_{GL}$, where $\alpha_4=7\zeta(3)(D/4\pi T)^2$ and $\rho_2=-2\alpha_4/\xi^2_{GL}$ are the $q^4$ coefficient of the static kernel and the $q^2$ correction to the friction [Eqs.~\eqref{eq:evenquartic} and \eqref{eq:w2}]. For any quantity that involves an integral over all pair momenta, $\bm q\to\bm q+q_0\hat{\bm n}$ is a change of the integration variable, so that in the strong-DP limit such a quantity has no nonreciprocal part to first order in $h$. The first-order nonreciprocity of the Gaussian sector is then contained entirely in the helical shift: it can be observed in measurements sensitive to the pair momentum itself (the helical modulation $q_0$, Josephson interferometry), but it does not appear in transport coefficients and noise. At weak DP, where $\mathcal K\propto-\psi^{(2)}(\tfrac12+x)$ is not proportional to $\partial_xA$, the shift removes only part of the odd terms and a finite nonreciprocal response remains. We refer to this as the removability property of the strong-DP limit; its consequences for the diode effect and for the magnetochiral anisotropy are discussed in Secs.~\ref{sec:SDE} and \ref{sec:MCA}.

\subsection{Helical state and the crossover function}
\label{sec:helix}

\begin{figure}[t]
\centering
\begin{tikzpicture}
\begin{semilogxaxis}[width=0.72\textwidth,height=7cm,
xlabel={$g=\Gamma_t/4\pi T$},
ylabel={$\mathfrak g(g)=q_0\,v_F^2/4\alpha_R h$},
ymin=0,ymax=1.08,
grid=both, grid style={gray!20},
legend pos=north west, legend style={font=\footnotesize}]
\addplot[blue,very thick] coordinates {
(0.005012,0.008464) (0.00646,0.01088) (0.008327,0.01397) (0.01073,0.01793)
(0.01384,0.02298) (0.01784,0.0294) (0.02299,0.03754) (0.02963,0.0478)
(0.0382,0.06067) (0.04924,0.07669) (0.06347,0.09645) (0.08181,0.1205)
(0.1055,0.1495) (0.1359,0.1839) (0.1752,0.2237) (0.2259,0.2691)
(0.2911,0.3193) (0.3753,0.3736) (0.4837,0.4306) (0.6236,0.4888)
(0.8038,0.5465) (1.036,0.6023) (1.335,0.6549) (1.721,0.7034)
(2.219,0.7472) (2.86,0.7861) (3.687,0.8202) (4.753,0.8497)
(6.126,0.8751) (7.897,0.8966) (10.18,0.9147) (13.12,0.9299)
(16.91,0.9426) (21.8,0.9531) (28.1,0.9618) (36.22,0.9689)
(46.69,0.9748) (60.18,0.9796) (77.58,0.9835) (100,0.9867)};
\addlegendentry{$\mathfrak g(g)$, Eq.~\eqref{eq:gfun}}
\addplot[red,dashed,thick,domain=0.005:0.42] {14*1.2020569*x/9.8696044};
\addlegendentry{$14\zeta(3)\,g/\pi^2$ (weak DP)}
\addplot[black,dotted,domain=0.005:100] {1};
\addlegendentry{universal limit}
\end{semilogxaxis}
\end{tikzpicture}
\caption{Disorder crossover of the helical modulation vector, Eq.~\eqref{eq:gfun}. For weak Dyakonov-Perel relaxation ($g\ll1$) the helix is suppressed, $q_0\propto(\alpha_R p_F)^2\tau\,\alpha_R h/Tv_F^2$; for $g\gg1$ it saturates at the universal, disorder-independent value $q_0=4\alpha_R h/v_F^2$. The same function (and its derivatives) controls the crossover of every entry of Table~\ref{tab:couplings}.}
\label{fig:crossover}
\end{figure}
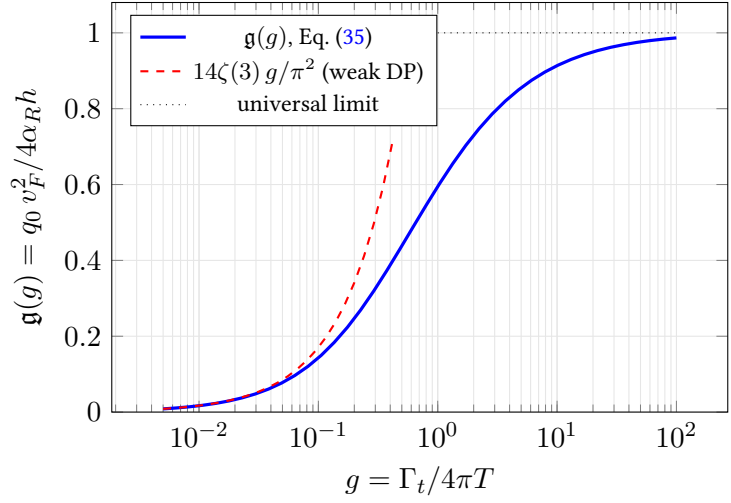

Minimizing $-L_R^{-1}$ at $\omega=0$ over the pair momentum shifts the ground state Cooper pair momentum to $\bm q_0=q_0\hat{\bm n}$ with
\begin{equation}
q_0=\frac{4\alpha_R h}{v_F^2}\;\mathfrak g\Big(\frac{\Gamma_t}{4\pi T}\Big),
\qquad
\mathfrak g(g)=1-\frac{\psi\big(\frac12+g\big)-\psi\big(\frac12\big)}{g\,\psi'\big(\frac12\big)} ,
\label{eq:gfun}
\end{equation}
shown in Fig.~\ref{fig:crossover}. The function $\mathfrak g$ interpolates between $\mathfrak g\simeq(14\zeta(3)/\pi^2)\,g$ at $g\ll1$ and $\mathfrak g\to1$ at $g\gg1$. Correspondingly, the helical wave vector grows linearly with the DP rate (hence with the scattering time, $q_0\propto(\alpha_R p_F)^2\tau\,\alpha_R h/Tv_F^2$) when $\Gamma_t\lesssim4\pi T$, and then saturates at the universal value $q_0= 4\alpha_R h/v_F^2$, which contains neither $\tau$ nor, apart from the trivial $\alpha_R h$ factor, the spin-orbit strength. This saturation is consistent with the known diffusive results for the anomalous phase gradient in Rashba systems \cite{BergeretTokatly:2014,HouzetMeyer:2015,VBT:2022}, while the weak-DP side reproduces expansions carried out at fixed order in $\alpha_R p_F\tau$. We stress that this resolves some earlier conflicting results about both the magnitude and the disorder dependence of the magnetoelectric couplings of the dirty Rashba model \cite{IlicBergeret:2024,VBT:2022}: Eq.~\eqref{eq:gfun} shows that both are correct within their domains, being the two tails of one crossover whose scale, $\Gamma_t\sim4\pi T$, i.e., $\alpha_R p_F\sim\sqrt{2\pi T/\tau}$, is parametrically small compared with the naive $\alpha_R p_F\tau\sim1$ boundary. In terms relevant to experiment, since $T\tau\ll1$ the relaxation-dominated regime is reached already for $\alpha_Rp_F\tau\gtrsim\sqrt{2\pi T\tau}$, which is still compatible with $\alpha_Rp_F\tau\ll1$; many Rashba superconductors are therefore in the universal regime near $T_c$, and the $(\alpha_R p_F\tau)^2$-suppressed formulas apply to systems with very weak spin-orbit coupling or strong disorder, or at elevated temperatures.

\subsection{Quartic sector}
\label{sec:quartic}

The uniform $|\Delta|^4$ coefficient follows from the third-order expansion of $\Tr[\Delta Q]$ in the cooperon generators, evaluated at the Gaussian saddle point (\ref{app:quartic}), and is the familiar disorder-independent result $b_0=7\zeta(3)/8\pi^2T^2$, in agreement with the Anderson theorem.

The momentum dependence of the quartic vertex is considerably more delicate than that of the quadratic kernel. The full vertex is a convolution over three independent pair momenta, and for the single-harmonic (plane-wave) states that govern the helical condensation and the depairing current the physical coefficient $b(\bm q)$, defined by $F_4=\tfrac{\nu}{2}\,b(\bm q)\,|\Delta|^4$ on the configuration $\Delta=|\Delta|e^{i\bm q\cdot\bm r}$, is not obtained by simply evaluating the single-mode kernel at the shifted cooperon pole: the internal structure of the vertex (in diagrammatic language, ladder rungs connecting nonadjacent sections of the loop) contributes to the gradient corrections at the same order. The even part of $b(\bm q)$ can be obtained exactly from the nonlinear saddle point of the sigma model, the Usadel equation, for which the plane-wave state is a standard pair-breaking problem. Expanding the Usadel free energy (\ref{app:quartic}) gives the per-frequency quartic weight $\omega_n/(\omega_n+Dq^2/2)^4$ rather than the naive $(\omega_n+Dq^2/2)^{-3}$, and hence a definite, and larger by the factor $4/3$, gradient correction. The momentum-odd part requires in addition the triplet-assisted corrections to the vertex at first order in the composite coupling $h\kappa_0$; the complete evaluation of all quartic diagrams, at weak spin-orbit coupling but arbitrary disorder, was carried out in Ref.~\cite{Hasan:2025}, whose diffusive limit we adopt here and check at the level of the diode observable in Sec.~\ref{sec:SDE}. That calculation refers to the weak-relaxation regime $\Gamma_t\ll4\pi T$; the crossover of the quartic invariants with $\Gamma_t/T$ is not computed in this paper, so the single-kernel description of the crossover applies to the quadratic sector, i.e., to $\bm\Lambda$, $\Xi$, $\bm\rho$, and $q_0$. Altogether, to the order needed below,
\begin{align}
&b(\bm q)=b_0\big[1+\bm{\mathfrak b}\cdot\bm q-\mathfrak b_2\,q^2\big],
\qquad
b_0=\frac{7\zeta(3)}{8\pi^2T^2}, \nonumber 
\\
&\bm{\mathfrak b}=\frac{124\,\zeta(5)}{7\,\zeta(3)}\,
\frac{(\alpha_R p_F\tau)^2}{\pi^2T^2}\;\bm n ,
\qquad
\mathfrak b_2=\frac{\pi^3}{42\,\zeta(3)}\,\frac{D}{T} .
\label{eq:bq}
\end{align}
The odd part carries the same $(\alpha_R p_F\tau)^2\,\bm n$ structure as all other invariants, with the dimensionless amplitude $124\zeta(5)/7\zeta(3)\pi^2\simeq1.55$. We emphasize that evaluating the uniform kernel at the shifted pole underestimates $\bm{\mathfrak b}$ by the factor $8/3$ (and the even coefficient $\mathfrak b_2$ by $4/3$), an error that would change the sign of the diode effect (Sec.~\ref{sec:SDE}). This can be seen explicitly in the even sector, where the Usadel equation provides the complete answer.

\subsection{The free energy}
\label{sec:freeenergy}

Collecting the static sector [$\omega=0$ terms of \eqref{eq:LRweak} and \eqref{eq:bq}] and restoring electromagnetic gauge invariance through $\bm\partial=-i\nabla-2e\bm A$ (all Lifshitz couplings descend from the same covariant derivative), the free energy density reads
\begin{align}
\frac{F[\Delta]}{\nu}=&
\Big(\frac{T-T_c}{T}+\frac{7\zeta(3)h^2}{4\pi^2T^2}\Big)|\Delta|^2
+\frac{\pi D}{8T}\,|\bm\partial\Delta|^2
-\bm\Lambda\cdot\Rey\big[\Delta^*\bm\partial\Delta\big]
\nonumber\\[4pt]
&+\Xi\;\Rey\big[(\bm\partial^2\Delta)^*\,(\hat{\bm n}\cdot\bm\partial)\Delta\big]+\frac{b_0}{2}\,|\Delta|^4
+\frac{b_0}{2}\,\bm{\mathfrak b}\cdot\Rey\big[|\Delta|^2\,\Delta^*\bm\partial\Delta\big].
\label{eq:GLfunctional}
\end{align}
with coefficients in Table~\ref{tab:couplings} (or, at arbitrary $\Gamma_t/T$, with the coefficients generated by the kernel \eqref{eq:Kkernel}). Three of its terms are odd in gradients: the linear invariant $\bm\Lambda$, the cubic invariant $\Xi$, and the quartic invariant $\bm{\mathfrak b}$. Their status is different. The $\bm\Lambda$ term can be removed from the quadratic sector by the substitution $\Delta\to\Delta\,e^{i\bm q_0\cdot\bm r}$ (corresponding to the helical state) and therefore drops out of any observable that probes only the quadratic form (we check this explicitly in Secs.~\ref{sec:SDE} and \ref{sec:MCA}). The invariants $\Xi$ and $\bm{\mathfrak b}$ survive the shift (up to corrections quadratic in the odd couplings) and thus determine the nonreciprocal observables, with one important refinement, worked out in Sec.~\ref{sec:SDE}: because the shift $\bm q_0$ is itself odd in $\bm h$, the even higher-gradient corrections to $a(\bm q)$ and $b(\bm q)$ [the $q^4$ term of the pair-breaking kernel and the $\mathfrak b_2$ term of Eq.~\eqref{eq:bq}] also generate odd terms of first order in $\alpha_R h$ and must be retained alongside $\Xi$ and $\bm{\mathfrak b}$ in any observable referenced to the helical ground state.

%==============================================================================================================================================================================================
%==============================================================================================================================================================================================
%==============================================================================================================================================================================================
%==============================================================================================================================================================================================
%==============================================================================================================================================================================================
\section{Dynamics: nonreciprocal friction, noise, and the stochastic TDGL equation}
\label{sec:TDGL}

\subsection{Nonreciprocal friction and the Onsager constraint}
\label{sec:friction}

We now turn to the dynamical content of Eq.~\eqref{eq:LRmaster}, which is inaccessible to free-energy methods. The dissipative dynamics of the order parameter is encoded in $\Imy L_R^{-1}$, which at low frequency defines the kinetic coefficient (friction) $\gamma(\bm q)$ of the pair field through $\Imy L_R^{-1}(\bm q,\omega)=\omega\gamma(\bm q)$. At $h=\alpha_R=0$ the familiar result is $\Imy L_R^{-1}=\pi\omega/8T$, i.e., $\gamma=\pi/8T$, the relaxation (overdamped) dynamics of the dirty-limit TDGL theory. In our system, expanding the magnetoelectric term of \eqref{eq:LRmaster} to linear order in $\omega$ (the static Lifshitz terms are real; the $\omega$ dependence enters through $x=(Dq^2-i\omega)/4\pi T$ in the argument of $\mathcal K$),
\begin{equation}
\Imy\,L_R^{-1}(\bm q,\omega)
=\omega\,\gamma(\bm q)=\frac{\pi\omega}{8T}\Big[1+\bm\rho\cdot\bm q\Big],
\qquad
\bm\rho\cdot\bm q
=-\frac{32}{\pi^2}\,h\kappa_0q_y\;\partial_x\mathcal K(x;\Gamma_t)\Big|_{\omega=0}
\label{eq:ImLR}
\end{equation}
where the weak-DP value, $\bm\rho=\frac{(\alpha_R p_F\tau)^2}{2T^2}\,\bm n$, follows from $\partial_x\mathcal K\to-\pi^2/32T^2$, and the strong-DP limit is listed in Table~\ref{tab:couplings}. Physically, a fluctuation of the pair field with momentum along $+\bm n$ decays at a different rate than one moving along $-\bm n$, so the lifetime of fluctuating Cooper pairs is nonreciprocal.

For the applications of Sec.~\ref{sec:apps} we shall also need the $q^2$ correction to the reciprocal part of the friction, which we denote by $\rho_2$. Expanding the pair-breaking term of \eqref{eq:LRmaster} to linear order in $\omega$ and to next order in $q^2$ gives $\Imy L_R^{-1}=(\pi\omega/8T)\big[1+\bm\rho\cdot\bm q+\rho_2q^2+\dots\big]$ with
\begin{equation}
\rho_2=\frac{\psi^{(2)}(\tfrac12)}{\psi'(\tfrac12)}\,\frac{D}{4\pi T}
=-\frac{7\zeta(3)}{\pi^3}\,\frac{D}{T}
=-\frac{2\alpha_4}{\xi^2_{GL}},
\qquad
\alpha_4=7\zeta(3)\Big(\frac{D}{4\pi T}\Big)^{\!2},
\label{eq:w2}
\end{equation}
where $\alpha_4$ is the $q^4$ coefficient of the static kernel [cf. Eq.~\eqref{eq:evenquartic}]. Both are field independent and, like $\xi^2_{GL}$, unaffected by the DP crossover; they generate odd terms after the helical shift, as in the diode problem.

Why is such a term allowed at all? Kinetic coefficients are constrained by Onsager reciprocity, which descends from microscopic time reversibility. The TDGL kinetic coefficient $\gamma(\bm q)$ connects the conjugate pair of variables $(\Delta_{\bm q},\Delta^*_{-\bm q})$. Under time reversal, momenta and the Zeeman field are odd, while the Rashba vector $\bm\alpha$ is even. The Onsager relation for the corresponding kinetic coefficient therefore reads
\begin{equation}
\gamma(\bm q;\bm h,\bm\alpha)=\gamma(-\bm q;-\bm h,\bm\alpha).
\label{eq:onsager}
\end{equation}
A momentum-odd contribution to $\gamma$ is thus permitted if and only if it is simultaneously odd in $\bm h$, as is the case for the microscopic result \eqref{eq:ImLR}, $\bm\rho\propto[\bm h\times\bm\alpha]$. On the other hand, a $\bm q$-odd, $\bm h$-even term corresponding to dissipative nonreciprocity without time-reversal breaking is forbidden by \eqref{eq:onsager} and does not appear in the calculation. Note that the positivity of dissipation requires $|\bm\rho\cdot\bm q|<1$, and since $\rho\sim(\alpha_R p_F\tau)^2\alpha_R h/T^2$ and $q\lesssim\xi_{GL}^{-1}$ within the GL description, this condition is parametrically well satisfied.

\subsection{Noise: the generalized fluctuation-dissipation relation}
\label{sec:noise}

The exact equilibrium relation \eqref{eq:FDT} has a further interesting consequence: it connects the nonreciprocal friction  to a nonreciprocal noise. The Keldysh block of the action \eqref{eq:SGL} is quadratic in the quantum component $\Delta^q$, and enters the weight $e^{iS}$ as $\exp[-2\nu\,\Delta^{*q}(-iL_K^{-1})\Delta^q]$ with $-iL_K^{-1}=2\coth(\omega/2T)\,\Imy L_R^{-1}>0$, a convergent Gaussian. Following the standard route \cite{LevchenkoKamenev:2007} we decouple it with a complex auxiliary field $\xi_\Delta$,
\begin{equation}
\exp\Big[-2\nu\,\Delta^{*q}\,\big({-i}L_K^{-1}\big)\,\Delta^{q}\Big]
=\int\mathcal D[\xi_\Delta]\,
\exp\Big[-\frac{2\nu\,|\xi_\Delta|^2}{-i L_K^{-1}}
+2i\nu\big(\Delta^{*q}\xi_\Delta+\xi^*_\Delta\Delta^{q}\big)\Big],
\end{equation}
after which the action is linear in $\Delta^{*q}$. Integrating over it enforces a delta-functional, $L_R^{-1}\Delta^c+\xi_\Delta=0$, and yields the stochastic equation of motion for $\Delta^{c}$, with $\xi_\Delta$ acting as a Langevin force whose correlator, $\langle\xi_\Delta\xi^*_\Delta\rangle=-iL_K^{-1}/2\nu$, is read off the Gaussian weight:
\begin{equation}
\big\langle\xi_\Delta(\bm q,\omega)\,\xi^*_\Delta(\bm q,\omega)\big\rangle
=\frac{\pi\omega}{8T\nu}\coth\Big(\frac{\omega}{2T}\Big)\big[1+\bm\rho\cdot\bm q\big]
\;\;\xrightarrow[\;\omega\ll T\;]{}\;\;
\frac{\pi}{4\nu}\big[1+\bm\rho\cdot\bm q\big],
\label{eq:noisefinal}
\end{equation}
together with $\langle\xi_\Delta\xi_\Delta\rangle=0$. This is the Langevin force of Eq.~\eqref{eq:TDGL} below, in which the kinetic coefficient $\gamma(\bm q)=(\pi/8T)(1+\bm\rho\cdot\bm q)$ multiplies $\partial_t\Delta$; the low-frequency intensity $(\pi/4\nu)(1+\bm\rho\cdot\bm q)=2T\gamma(\bm q)/\nu$ is the Einstein relation for the free energy $F=\nu a|\Delta|^2$, and we check below that it reproduces the equilibrium amplitude \eqref{eq:gibbs}. \footnote{Ref.~\cite{LevchenkoKamenev:2007} writes the TDGL equation divided by the bare kinetic coefficient $\pi/8T$, so that the correlator quoted there is larger by $(8T/\pi)^2$.} In real space the equal-time correlator acquires an anti-Hermitian gradient correction,
\begin{equation}
\big\langle\xi_\Delta(\bm r,t)\,\xi^*_\Delta(\bm r',t')\big\rangle
=\frac{\pi}{4\nu}\,\delta(t-t')
\Big[\delta(\bm r-\bm r')-i\,\bm\rho\cdot\nabla_{\bm r}\,\delta(\bm r-\bm r')\Big],
\label{eq:noisereal}
\end{equation}
which is Hermitian as an integral kernel, $K(\bm r,\bm r')=K^*(\bm r',\bm r)$, and positive definite for $\rho q<1$. The noise experienced by pair fluctuations propagating along $+\bm n$ is stronger than for those propagating along $-\bm n$, and at finite frequency the fluctuation spectrum of the order parameter is nonreciprocal, $\langle|\Delta_{\bm q}|^2\rangle_\omega\neq\langle|\Delta_{-\bm q}|^2\rangle_\omega$.

We make two further observations. (i) In equilibrium there are no independent nonreciprocal noise constants. The momentum-odd noise \eqref{eq:noisefinal} goes hand-in-hand with the momentum-odd friction \eqref{eq:ImLR} by the exact identity \eqref{eq:FDT}. A phenomenological TDGL model that includes a drift-like term in the dynamics but keeps isotropic white noise violates the FDT at order $\bm\rho\cdot\bm q$ and produces spurious equilibrium currents. (ii) Equal-time correlations are insensitive to the kinetic invariant. Solving the linear Langevin dynamics defined by Eqs.~\eqref{eq:TDGL} and \eqref{eq:noisefinal} mode by mode (an Ornstein--Uhlenbeck problem with friction $\gamma(\bm q)=\tfrac{\pi}{8T}(1+\bm\rho\cdot\bm q)$ and noise power $\propto(1+\bm\rho\cdot\bm q)$) the stationary equal-time correlator comes out as
\begin{equation}
\big\langle|\Delta_{\bm q}|^2\big\rangle=\frac{T}{\nu\,a(\bm q)},
\qquad
a(\bm q)=-L_R^{-1}(\bm q,0),
\label{eq:gibbs}
\end{equation}
with the two factors of $(1+\bm\rho\cdot\bm q)$ canceling. This is an Einstein relation: equilibrium thermodynamics is insensitive to kinetic coefficients. The equal-time spectrum \eqref{eq:gibbs} is nonreciprocal, $\langle|\Delta_{\bm q}|^2\rangle\neq\langle|\Delta_{-\bm q}|^2\rangle$, but only through the thermodynamic invariants $\bm\Lambda$ and $\Xi$ contained in $a(\bm q)$; the kinetic invariant does not enter it. As a corollary the equilibrium supercurrent vanishes identically, as one also sees directly: the current $\int_{\bm q}a'(\bm q)\,T/a(\bm q)$ is the integral of a total derivative. The kinetic invariant thus affects only the dynamics of relaxation, the finite-frequency noise spectra, and the nonlinear response.

\subsection{The stochastic TDGL equation}
\label{sec:tdgleq}

Assembling the classical saddle point of the effective action with the noise \eqref{eq:noisefinal}, and using the covariant momentum $\bm\partial=-i\nabla-2e\bm A$ of Sec.~\ref{sec:freeenergy}, the stochastic TDGL equation of a disordered Rashba superconductor takes the form
\begin{equation}
\begin{aligned}
\frac{\pi}{8T}\big(1+\,\bm\rho\cdot\bm\partial\big)\,\partial_t\Delta
=\Big[
\frac{T_c-T}{T}&-\frac{7\zeta(3)h^2}{4\pi^2T^2}
-\frac{\pi D}{8T}\,\bm\partial^2
+\,\bm\Lambda\cdot\bm\partial
\\[2pt]
&-\,\Xi\,(\hat{\bm n}\cdot\bm\partial)\,\bm\partial^2
\Big]\Delta
-\frac1\nu\frac{\delta F_4}{\delta\Delta^*}
+\xi_\Delta(\bm r,t) ,
\end{aligned}
\label{eq:TDGL}
\end{equation}
where the quartic term is the functional derivative of $F_4$, the last two terms of Eq.~\eqref{eq:GLfunctional},
\begin{equation}
-\frac1\nu\frac{\delta F_4}{\delta\Delta^*}
=-b_0|\Delta|^2\Delta-\frac{b_0}{4}\,\bm{\mathfrak b}\cdot\Big[2|\Delta|^2\bm\partial\Delta+\Delta^2(\bm\partial\Delta)^*+\bm\partial\big(|\Delta|^2\Delta\big)\Big].
\label{eq:F4force}
\end{equation}
For $\Delta\propto e^{i\bm q\cdot\bm r}$ one has $\bm\partial\to\bm q$, so that the static bracket is $L_R^{-1}(\bm q,0)=-a(\bm q)$ of Eq.~\eqref{eq:LRweak}, the left-hand side is $\gamma(\bm q)\partial_t\Delta$, and the quartic term reduces to $-b(\bm q)|\Delta|^2\Delta$ with $b(\bm q)$ of Eq.~\eqref{eq:bq}. For a spatially varying order parameter the gradient inside the quartic term must be placed as dictated by the functional derivative of $F_4$ in Eq.~\eqref{eq:GLfunctional}. The even gradient correction $\mathfrak b_2$ of Eq.~\eqref{eq:bq} enters in the same way, and its action on a general $\Delta(\bm r)$ would require the full three-momentum structure of the vertex, which we have not computed. Equation \eqref{eq:TDGL} contains four distinct nonreciprocal couplings, all proportional to $(\alpha_R p_F\tau)^2\bm n$ in the weak-DP regime (Table~\ref{tab:couplings}) but with different physical meaning:  the linear Lifshitz invariant $\bm\Lambda$, removable by a momentum shift; the cubic invariant $\Xi$ which is the main source of the diode effect; $\bm{\mathfrak b}$ inside $F_4$, a second source of the diode effect, acting alongside the even higher-gradient terms as explained in Sec.~\ref{sec:SDE}; and the new kinetic invariant  $\bm\rho$ that gives rise to the nonreciprocal relaxation and noise, invisible in statics but active in nonlinear transport. To our knowledge, Eq.~\eqref{eq:TDGL} with the correlator \eqref{eq:noisefinal} is the first stochastic TDGL theory of a noncentrosymmetric superconductor derived from a microscopic model.

Two remarks concern the use of Eq.~\eqref{eq:TDGL} in an applied electric field, which is needed in Sec.~\ref{sec:MCA}. First, in a uniform field introduced through $\bm A=-\bm Et$ the covariant momentum of a mode with canonical momentum $\bm q$ drifts, $\bm\partial\to\bm k(t)=\bm q+2e\bm Et$, so that all coefficients of \eqref{eq:TDGL} become functions of the kinetic momentum at the time at which they act. Because the kinetic momentum and the time derivative do not commute in a field, $[\partial_t,\bm\partial]=2e\bm E$, gauge covariance alone does not fix the mixed term on the left-hand side: $(1+\bm\rho\cdot\bm\partial)\partial_t\Delta$ and $\partial_t[(1+\bm\rho\cdot\bm\partial)\Delta]$ differ by $2e\bm\rho\cdot\bm E\,\Delta$. The ordering written in \eqref{eq:TDGL}, with the kinetic coefficient evaluated at the instantaneous momentum and acting on $\partial_t\Delta$, is the one for which the friction and the noise \eqref{eq:noisefinal} satisfy the Einstein relation at every instant, so that each mode relaxes toward the Gibbs occupation \eqref{eq:gibbs} of its instantaneous momentum. This is the condition of local detailed balance. It extends the equilibrium fluctuation-dissipation relation \eqref{eq:FDT} to the driven state, and it is required for the driven dynamics to obey the fluctuation theorem. In \ref{app:MCA} we show that \eqref{eq:TDGL} satisfies the fluctuation theorem while the alternative orderings violate it. The choice of the ordering is important: with the symmetrized ordering the fluctuation magnetochiral anisotropy of Sec.~\ref{sec:MCA} would vanish, because $\rho=\Xi/\xi^2_{GL}$ in the Rashba model. Second, the electric current is obtained from the variation of the action with respect to the quantum component of the vector potential. The static part of the kernel gives the thermodynamic vertex $2e\nu\,\partial_{\bm q}a(\bm q)\,|\Delta_{\bm q}|^2$, while the dissipative term contributes vertices of opposite sign in the retarded and advanced blocks, which cancel in the classical current (\ref{app:MCA}); the current carried by the fluctuations is therefore the thermodynamic one, as used in Sec.~\ref{sec:MCA}.

%======================================================================
\section{Applications}
\label{sec:apps}

\subsection{Superconducting diode effect near $T_c$}
\label{sec:SDE}

As a first application, consider the nonreciprocity of the depairing (critical) current just below $T_c$. We look for uniform current-carrying solutions $\Delta=|\Delta|e^{iq r_n}$ with the phase winding along the nonreciprocal axis $\hat{\bm n}$. This is the direction of maximal nonreciprocity. In the in-plane field the response is anisotropic: for a current at an angle $\theta$ to $\hat{\bm n}$ all odd couplings enter through $\hat{\bm n}\cdot\bm q=q\cos\theta$, so that the diode efficiency computed below is multiplied by $\cos\theta$ and vanishes for a current along the field, while the reciprocal coefficients remain isotropic in the plane to the order we work (the only anisotropic even terms generated by the kernel \eqref{eq:Ainv} are of order $(\kappa_0q)^2\sim\alpha_R^6$). The same angular dependence holds for the magnetochiral anisotropy of Sec.~\ref{sec:MCA}.

One important nuance must be clarified before formulating the free energy: for the diode coefficient the gradient expansion has to be carried one order beyond what is displayed in Eq.~\eqref{eq:GLfunctional}. The critical currents are reached at $\bar q\sim\sqrt{|\epsilon|}/\xi_{GL}$ measured from the helical minimum $q_0\propto\alpha_R h$; since the shift of the expansion point is itself odd in $\bm h$, every even coefficient of the theory spawns odd-in-$\bar q$ terms of first order in $\alpha_R h$. Powers of the momentum and powers of the nonreciprocal coupling must therefore be counted independently: to obtain $\eta$ at first order in $\alpha_R h$ and leading order in $|\epsilon|$ one needs $a(q)$ through $q^4$ and $b(q)$ through $q^2$ \cite{Hasan:2024}. Keeping only $\Xi$ and $\mathfrak b$ is inconsistent and, as we shall see, gets even the sign of the effect wrong. From Eq.~\eqref{eq:GLfunctional}, supplemented by the next order of the expansion of the pair-breaking kernel of \eqref{eq:LRweak} [the $x^2$ term of $\psi(\tfrac12+x)$, $x=Dq^2/4\pi T$] and by the quartic-vertex gradient correction of Eq.~\eqref{eq:bq}, the condensation energy at fixed $q$ is governed by
\begin{equation}
a(q)=\epsilon-\Lambda q+\xi_{GL}^2\,q^2+\Xi\,q^3-\alpha_4 q^4,
\qquad
b(q)=b_0\big(1+\mathfrak b\,q-\mathfrak b_2\,q^2\big),
\end{equation}
with
\begin{equation}
\alpha_4=7\zeta(3)\Big(\frac{D}{4\pi T}\Big)^{\!2}
=\frac{28\,\zeta(3)}{\pi^4}\,\xi_{GL}^4 ,
\qquad
\mathfrak b_2=\frac{\pi^3}{42\,\zeta(3)}\frac{D}{T}
=\frac{4\pi^2}{21\,\zeta(3)}\,\xi_{GL}^2 ,
\label{eq:evenquartic}
\end{equation}
and $\Lambda=|\bm\Lambda|$, $\mathfrak b=|\bm{\mathfrak b}|$. Both $\alpha_4$ and $\mathfrak b_2$ are even, field-independent coefficients; they are pure numbers times the appropriate powers of $\xi_{GL}$, which is why they enter the diode coefficient with $O(1)$ weights. For $a(q)<0$ the amplitude adjusts to $|\Delta|^2=-a(q)/b(q)$ and the supercurrent follows from differentiating the free energy with respect to the vector potential, i.e., with respect to $q$ at fixed $\Delta$:
\begin{equation}
j(q)=2e\nu\Big[a'(q)\,|\Delta|^2+\tfrac12\,b'(q)\,|\Delta|^4\Big]
=-\frac{2e\nu}{b(q)}\,a(q)\,a'(q)+\frac{e\nu\,b'(q)}{b^2(q)}\,a^2(q) .
\label{eq:jq}
\end{equation}
Both terms matter: the second, often omitted, contribution of the $q$-dependent quartic coefficient enters the diode asymmetry at the same order as the first. The critical currents $j_c^\pm$ are the extrema of $j(q)$ on the two branches.

A single reduced function provides the most transparent way to keep track of the various contributions
\begin{equation}
\tilde a(q)\equiv a(q)\sqrt{\frac{b_0}{b(q)}},
\qquad
\frac{F_{\rm cond}}{\nu}=-\frac{a^2(q)}{2b(q)}=-\frac{\tilde a^2(q)}{2b_0},
\qquad
j(q)=-\frac{2e\nu}{b_0}\,\tilde a(q)\,\tilde a'(q),
\end{equation}
which reduces the problem to one with a momentum-independent quartic coefficient \cite{Hasan:2024}. Expanding $\sqrt{b_0/b}=1-\tfrac12\mathfrak b\,q+\tfrac12\mathfrak b_2\,q^2+\dots$ and keeping all terms of first order in the odd couplings,
\begin{equation}
\tilde a(q)=\epsilon
-\Big(\Lambda+\frac{\epsilon\,\mathfrak b}{2}\Big)q
+\Big(\xi_{GL}^2+\frac{\epsilon\,\mathfrak b_2}{2}\Big)q^2
+\Big(\Xi-\frac{\xi^2_{GL}\mathfrak b}{2}-\frac{\Lambda\,\mathfrak b_2}{2}\Big)q^3
-\Big(\alpha_4-\frac{\xi^2_{GL}\mathfrak b_2}{2}\Big)q^4 .
\end{equation}
Note the cross terms: the odd vertex $\mathfrak b$ mixes with $\epsilon$ and $\xi^2_{GL}$, and the even vertex correction $\mathfrak b_2$ mixes with the odd couplings. The linear term is removed by the shift $q\to q_0+\bar q$ with $q_0=\Lambda/2\xi_{GL}^2$ [up to corrections of $O(\epsilon)$ and $O(\alpha_R h)^2$ that do not affect the leading asymmetry], which leaves
\begin{equation}
\tilde a(\bar q)=\bar\epsilon+\xi^2_{GL}\,\bar q^{\,2}
+a_3\,\bar q^{\,3}
-\alpha_4\bar q^{\,4}+\dots,
\qquad
\bar\epsilon=\epsilon-\frac{\Lambda^2}{4\xi^2_{GL}}<0 ,
\end{equation}
so that $\Lambda$ survives as a shift of $T_c$ and through the complete odd cubic coefficient
\begin{equation}
a_3=\Xi-\frac{\xi^2_{GL}\mathfrak b}{2}
-\frac{\Lambda\,\mathfrak b_2}{2}
-4\Big(\alpha_4-\frac{\xi^2_{GL}\mathfrak b_2}{2}\Big)q_0
=\Xi-\frac{\xi^2_{GL}\mathfrak b}{2}
-\frac{2\Lambda\alpha_4}{\xi^2_{GL}}
+\frac{\Lambda\,\mathfrak b_2}{2} .
\label{eq:a3}
\end{equation}
The first two terms are the irreducible odd invariants; the third is the even $q^4$ coefficient promoted to an odd term by the helical shift ($-4\alpha_4q_0\bar q^3$ from expanding $-\alpha_4(q_0+\bar q)^4$); the last combines the two $\mathfrak b_2$ cross terms. Truncating $a(q)$ at $q^3$ and $b(q)$ at $q$ amounts to dropping the last two entries of \eqref{eq:a3}, which are of the same order. The extrema of $j=-2e\nu\tilde a\tilde a'/b_0$ lie at $\bar q_c^{\pm}\simeq\pm\sqrt{|\bar\epsilon|/3\xi^2_{GL}}$, where $|j_0|=\frac{8e\nu}{3\sqrt3\,b_0}\,\xi_{GL}|\bar\epsilon|^{3/2}$, and to first order in $a_3$ the diode efficiency is $\eta=\sqrt{|\bar\epsilon|}\,a_3/\sqrt3\,\xi^3_{GL}$ \cite{Hasan:2024}. Collecting Table~\ref{tab:couplings} and Eqs.~\eqref{eq:bq} and \eqref{eq:evenquartic}, the four contributions to the bracket $2a_3/\xi^2_{GL}$, in units of $(\alpha_R p_F\tau)^2\alpha_R h/T^2$
give the diode efficiency
\begin{equation}
\begin{aligned}
\eta\equiv\frac{j_c^+-|j_c^-|}{j_c^++|j_c^-|}
&=\frac{\sqrt3}{6}\Big[\frac{2\Xi}{\xi^2_{GL}}-\mathfrak b
-\frac{4\Lambda\alpha_4}{\xi^4_{GL}}+\frac{\Lambda\mathfrak b_2}{\xi^2_{GL}}\Big]
\frac{\sqrt{|\bar\epsilon|}}{\xi_{GL}}
\\[4pt]
&\simeq-0.136\,\frac{(\alpha_R p_F\tau)^2\,\alpha_R h}{T^2}\,
\frac{\sqrt{|\bar\epsilon|}}{\xi_{GL}} .
\end{aligned}
\label{eq:eta}
\end{equation}
Several comments are in order. First, the scaling $\eta\propto\sqrt{T_c-T}$ is the generic GL result \cite{Daido:2022,IlicBergeret:2022,Hasan:2024}: the asymmetry is carried by the odd terms evaluated at $q_c\propto\sqrt{|\bar\epsilon|}$. Second, the linear invariant cancels from the quadratic sector, but it is not a spectator: it reenters the observable through the products $\Lambda\alpha_4$ and $\Lambda\mathfrak b_2$, 
since the helical shift it generates converts even gradient corrections into odd ones. Third, the contributing terms interfere: the cubic invariant ($+1$) is overcompensated by the quartic vertex and the $q^4$-gradient terms, with the net bracket $5/3-124\zeta(5)/7\zeta(3)\pi^2-392\zeta^2(3)/\pi^6\simeq-0.471$. The diode polarity is therefore opposite to what the cubic invariant alone (or the truncated pair $\Xi,\mathfrak b$ with the single-mode vertex) would predict, while the magnitude remains comparable. This sign reversal is a prediction specific to the weak spin-orbit, diffusive regime of the Rashba model. Fourth, the result \eqref{eq:eta} can be checked against the independent diagrammatic calculation of the same model \cite{Hasan:2025}, which evaluates all Ginzburg-Landau coefficients at weak spin-orbit coupling and arbitrary $T\tau$: substituting the diffusive limits of Eqs.~(C4a)--(C4c) of that work into the general efficiency formula of Ref.~\cite{Hasan:2024} and converting conventions, Eq.~\eqref{eq:eta} reproduces the diffusive-limit efficiency quoted there, $\eta=-0.307\,(T_c\tau)^{3/2}(h/T_c)(\alpha_R/v_F)(\alpha_R p_F/T_c)^2\sqrt{(T_c-T)/T_c}$; equivalently, the individual couplings $\Lambda$, $\Xi$, and $\alpha_4$ match term by term, while $\mathfrak b$ and $\mathfrak b_2$ are fixed by that comparison (Sec.~\ref{sec:quartic}). We have also checked the perturbative formula by numerical extremization of $j(q)$. Finally, in the strong-DP regime the same four-term formula holds with the couplings replaced by their crossover forms generated by the kernel \eqref{eq:Kkernel} (with the quartic-sector crossovers requiring the vertex analysis of \ref{app:quartic}). There the removability property of Sec.~\ref{sec:master} has a direct consequence: by the relation $\Xi=2\Lambda\alpha_4/\xi^2_{GL}$ of Eq.~\eqref{eq:removability} the two quadratic-sector terms of \eqref{eq:eta} cancel, $2\Xi/\xi^2_{GL}-4\Lambda\alpha_4/\xi^4_{GL}\to0$, so that at strong DP the diode effect is determined entirely by the remaining quartic-vertex terms $-\mathfrak b+\Lambda\mathfrak b_2/\xi^2_{GL}$, in accord with the disorder-driven diode effect found at strong spin-orbit coupling in Ref.~\cite{Hasan:2025}; since $\Lambda$, $\Xi$, and $\bm\rho$ retain their signs across the entire disorder crossover [see the discussion following Eq.~\eqref{eq:LRstrong}], any disorder-induced reversal of $\eta$, such as that found in the weak spin-orbit regime in Ref.~\cite{Hasan:2025}, originates from the quartic sector rather than from the cubic invariant itself. For realistic parameters, e.g. $\alpha_R p_F\tau=0.3$, $h/T=0.5$, $T\tau=0.1$, $E_F\tau=50$, and $\epsilon=0.1$ one concludes that $|\eta|\ll1$, so for weak spin-orbit the intrinsic diode effect of the bulk depairing current is small, which is consistent with the fact that sizable observed efficiencies typically involve either strong Rashba splitting or extrinsic mechanisms \cite{Nadeem:2023, Shaffer:2025}.

\subsection{Magnetochiral anisotropy from superconducting fluctuations}
\label{sec:MCA}

Above $T_c$ the order parameter exists only as a fluctuation, yet it dominates the temperature dependence of transport near the transition through the Aslamazov-Larkin (AL) paraconductivity \cite{AL:1968,LarkinVarlamov:2005}. In a noncentrosymmetric superconductor the fluctuation spectrum is tilted by the Lifshitz invariants and damped anisotropically by the kinetic invariant $\bm\rho$, so the paraconductivity must contain a nonreciprocal component, the fluctuation-induced magnetochiral anisotropy. The stochastic TDGL theory derived above is needed for its calculation: the noise maintains the fluctuation population whose field-induced redistribution carries the nonlinear current.

The calculation is transparent in the language of a kinetic equation for the mode occupations $n_{\bm q}=\langle|\Delta_{\bm q}|^2\rangle$. In a dc electric field, $\bm E=E\hat{\bm n}$ (for a general in-plane direction the rectified current below is multiplied by $\cos\theta$, Sec.~\ref{sec:SDE}), introduced through $\bm A=-\bm Et$, the kinetic pair momentum of each fluctuation drifts via $\dot{\bm q}=2e\bm E$, and the coefficients of Eq.~\eqref{eq:TDGL} are evaluated at the instantaneous momentum, as discussed in Sec.~\ref{sec:tdgleq}.  Relaxation (at the nonreciprocal rate $\lambda_{\bm q}$) and noise, meanwhile, drive $n_{\bm q}$ toward the instantaneous local-equilibrium value \eqref{eq:gibbs}. The stationary balance reads
\begin{equation}
2e\bm E\cdot\frac{\partial n_{\bm q}}{\partial\bm q}
=-2\lambda_{\bm q}\Big[n_{\bm q}-\frac{T}{\nu\,a(\bm q)}\Big],
\qquad
\lambda_{\bm q}=\frac{8T}{\pi}\,\frac{a(\bm q)}{\Gamma_\rho(\bm q)},
\quad
\Gamma_\rho(\bm q)=1+\bm\rho\cdot\bm q+\rho_2q^2+\dots,
\label{eq:kinetic}
\end{equation}
where the form of the right-hand side is dictated by the generalized FDT, Eq.~\eqref{eq:noisefinal}, and $\Gamma_\rho=\Imy L_R^{-1}/(\pi\omega/8T)$ is the full friction function of Sec.~\ref{sec:friction}.
The current carried by the fluctuations is $j=2e\nu\int\!\frac{d^2q}{(2\pi)^2}\,a'_n(\bm q)\,n_{\bm q}$, with $a'_n=\partial a/\partial q_n$ the velocity of the mode; the kinetic term of \eqref{eq:TDGL} does not contribute to the current (Sec.~\ref{sec:tdgleq}, \ref{app:MCA}).

The subtlety encountered in the diode problem arises here in a slightly different form. The rectified current is odd in $h$ and is obtained at first order in the odd couplings, and one might expect that only $\Lambda$, $\Xi$, and $\bm\rho$ can contribute, the even gradient corrections $-\alpha_4q^4$ in $a(\bm q)$ and $\rho_2q^2$ in $\Gamma_\rho(\bm q)$ being suppressed by a further power of $\epsilon$. This is not the case, because the contributions linear in $\Lambda$ are individually more singular, $\propto\Lambda/\epsilon^3$, than those of $\Xi$ and $\bm\rho$, $\propto1/\epsilon^2$. They cancel at that order (\ref{app:MCA}), but a single insertion of $\alpha_4$ or $\rho_2$ adds one power of $\epsilon$, so that the $\Lambda\alpha_4$ and $\Lambda\rho_2$ cross terms are of order $1/\epsilon^2$, the same as the contributions of $\bm\rho$ and $\Xi$. It is again convenient to measure all momenta from the helical minimum, $\bm q\to\bm q+q_0\hat{\bm n}$ with $q_0=\Lambda/2\xi^2_{GL}$. The kinetic equation \eqref{eq:kinetic} is form-invariant under this change of variables (the drive $\partial_{\bm q}$, the integration measure, and the current vertex $\partial_na$ are all unaffected), while $a(\bm q)$ and $\Gamma_\rho(\bm q)$ lose their linear terms and acquire the effective odd couplings
\begin{equation}
\rho_{\rm eff}=\rho+2\rho_2q_0=\rho-\frac{2\Lambda\alpha_4}{\xi^4_{GL}},
\qquad
\frac{\Xi_{\rm eff}}{\xi^2_{GL}}=\frac{\Xi-4\alpha_4q_0}{\xi^2_{GL}}=\frac{\Xi}{\xi^2_{GL}}-\frac{2\Lambda\alpha_4}{\xi^4_{GL}},
\label{eq:effcouplings}
\end{equation}
where $\rho_2=-2\alpha_4/\xi^2_{GL}$, Eq.~\eqref{eq:w2}, was used; the remaining even corrections contribute only at relative order $\epsilon$. The rest of the calculation is unchanged. Solving \eqref{eq:kinetic} iteratively in $E$ and collecting the terms odd under $\bm q\to-\bm q$ we obtain (details and all intermediate integrals in \ref{app:MCA}):
\begin{align}
&j=\sigma_{AL}E+\delta j,
\qquad
\sigma_{AL}=\frac{e^2}{16\,\epsilon},\nonumber
\\
&\delta j=\frac{\pi e^3E^2}{256\,T}\,
\frac{\rho_{\rm eff}+\Xi_{\rm eff}/\xi^2_{GL}}{\epsilon^{2}}
=\frac{\pi e^3E^2}{128\,T}\,R\Big(\frac{\Gamma_t}{4\pi T}\Big)\,
\frac{\rho}{\epsilon^{2}}.
\label{eq:MCAresult}
\end{align}
Here $\rho$ is the kinetic coupling of Eq.~\eqref{eq:ImLR} at the given $\Gamma_t/T$ (Table~\ref{tab:couplings}), we used the identity $\rho=\Xi/\xi^2_{GL}$, which holds at every $\Gamma_t/T$ because both couplings are proportional to $\partial_x\mathcal K(0;\Gamma_t)$ [Eqs.~\eqref{eq:ImLR} and the discussion after Eq.~\eqref{eq:LRstrong}], and introduced the irreducibility factor
\begin{equation}
R(g)\equiv\frac{\rho_{\rm eff}+\Xi_{\rm eff}/\xi^2_{GL}}{\rho+\Xi/\xi^2_{GL}}
=1-\frac{28\zeta(3)}{\pi^2}\,\frac{\mathcal K(0;\Gamma_t)}{-\partial_x\mathcal K(0;\Gamma_t)}
\simeq\begin{cases}
\displaystyle 0.411, &g\ll1,\\[10pt]
\displaystyle \frac{2}{\pi^2}\,\frac{\ln g}{g}, & g\gg1,
\end{cases}
\label{eq:Rfactor}
\end{equation}
with $g=\Gamma_t/4\pi T$, which measures the fraction of the first-order nonreciprocity that cannot be absorbed into the helical shift. In the weak-DP regime $R\simeq0.411$ and $\rho=(\alpha_Rp_F\tau)^2\alpha_Rh/2T^2$. The factor $1-392\zeta^2(3)/\pi^6\approx0.411$ is the same combination that appears in the diode coefficient of Eq.~\eqref{eq:eta}, since it originates from the same $\Lambda\alpha_4$ mechanism. The linear-response term reproduces the textbook AL conductivity of a two-dimensional film.
The nonlinear term exhibits four notable features. First, the linear invariant $\Lambda$ drops out of the leading, $1/\epsilon^3$, order: its four contributions to the rectification integrand cancel. It reenters at order $1/\epsilon^2$ only through the cross terms $\Lambda\alpha_4$ and $\Lambda \rho_2$ of Eq.~\eqref{eq:effcouplings}, as it reenters the diode coefficient through Eq.~\eqref{eq:a3}; both cross terms are evaluated in \ref{app:MCA} and sum to $-4\Lambda\alpha_4/\xi^4_{GL}$. Second, the kinetic invariant $\bm\rho$ and the cubic invariant $\Xi/\xi^2_{GL}$ contribute equally: in two dimensions their weights coincide, the bare couplings are equal, $\rho=\Xi/\xi^2_{GL}$, at every disorder strength, and the shift corrections reduce both by the same amount, so that the effective couplings remain equal across the whole crossover. The noise/friction nonreciprocity is thus not a spectator: it carries half of the fluctuation MCA. This equal division relies on the ordering of the kinetic term in an electric field discussed in Sec.~\ref{sec:tdgleq}; with the symmetrized ordering the two contributions would cancel instead of adding (\ref{app:MCA}). Third, $\delta j\propto\epsilon^{-2}$ diverges faster than $\sigma_{AL}\propto\epsilon^{-1}$: fluctuations amplify the magnetochiral anisotropy on approach to $T_c$, in qualitative agreement with the experimental observations of nonreciprocal transport enhancement near the transition \cite{Wakatsuki:2017,Itahashi:2020} and with fluctuation analyses of related models \cite{WakatsukiNagaosa:2018,Hoshino:2018,DeMiranda:2026}. Fourth, the amplitude does not follow the couplings of Table~\ref{tab:couplings}: beyond the DP crossover $\rho$ and $\Xi$ saturate, while $R$ decreases as $(\ln g)/g$, Eq.~\eqref{eq:Rfactor}, so that $\delta j$ is smaller by this factor than the estimate based on the saturated couplings. This is a consequence of the removability property of Sec.~\ref{sec:master}: at strong DP the first-order nonreciprocity of the Gaussian sector reduces to the helical shift, which does not contribute to a transport coefficient, since the latter involves an integral over all pair momenta. Using the strong-DP entries of Table~\ref{tab:couplings} in \eqref{eq:MCAresult} without the factor $R$ would therefore overestimate the effect.

The result acquires a transparent form when expressed through the current rather than the field. The fluctuation current adds to the normal (Drude) current of the film, $j=(\sigma_N+\sigma_{AL})E+\delta j$ with $\sigma_N=2e^2\nu D$, so that the nonreciprocal part of the resistance at fixed current is
\begin{equation}
\frac{\delta R}{R}=-\frac{\delta j}{(\sigma_N+\sigma_{AL})E},
\qquad E=\frac{j}{\sigma_N+\sigma_{AL}} .
\label{eq:MCAgeneral}
\end{equation}
Two limits are instructive. If the fluctuation conductivity dominates, $\sigma_{AL}\gg\sigma_N$,
\begin{equation}
\frac{\delta R}{R}=-\frac{\delta j}{\sigma_{AL}E}
=-R\Big(\frac{\Gamma_t}{4\pi T}\Big)\,\frac{2\pi\rho}{e\,T}\;j\;\big[1+O(\epsilon)\big]:
\label{eq:MCAj}
\end{equation}
the $\epsilon$ factors cancel between $\delta j$ and $\sigma_{AL}^2$, and at fixed current density the magnetochiral signal is independent of the distance to $T_c$. In the opposite limit, $\sigma_N\gg\sigma_{AL}$, Eq.~\eqref{eq:MCAgeneral} gives $\delta R/R=-\delta j/\sigma_NE=-(\pi e^3/128T)R\rho\,j/(\sigma_N^2\epsilon^2)$, which grows as $\epsilon^{-2}$ on approach to $T_c$. Which limit applies depends on the sheet conductance: $\sigma_{AL}/\sigma_N=1/(32\,\epsilon\,\nu D)$, while the Gaussian description used here requires $\epsilon\gg\mathrm{Gi}$ with the Ginzburg number of a two-dimensional film also of order $1/\nu D$ \cite{LarkinVarlamov:2005}. In a two-dimensional film the fluctuation conductivity therefore remains a correction to the Drude conductivity over most of the Gaussian regime, where the measured nonreciprocal resistance follows the $\epsilon^{-2}$ law, and the plateau \eqref{eq:MCAj} is reached only in films of low sheet conductance close to $T_c$, near the limit of applicability of the Gaussian approximation; the general expression \eqref{eq:MCAgeneral} interpolates between the two. The remaining conditions of validity are those of the calculation: the Gaussian regime $\epsilon\gg\mathrm{Gi}$, classical frequencies, and the weak-field expansion, which requires the drift of a fluctuation during its lifetime to be small compared with its characteristic momentum, $2eE\tau_{GL}\ll\sqrt\epsilon/\xi_{GL}$, i.e., $eE\ll4T\epsilon^{3/2}/\pi\xi_{GL}$. Under these conditions the temperature independence of $\gamma_{\rm MCA}=\delta R/(R\,h\,j)$ in the fluctuation-dominated limit, crossing over to the rapidly growing vortex and amplitude mechanisms below $T_c$ \cite{Hoshino:2018}, remains a distinctive signature of the mechanism described here, with a magnitude proportional to $R\rho$, Eq.~\eqref{eq:MCAj}.

%==============================================================================================================================================================================================
%==============================================================================================================================================================================================
%==============================================================================================================================================================================================
%==============================================================================================================================================================================================
%==============================================================================================================================================================================================
\section{Summary and outlook}
\label{sec:conclusions}

We have constructed the Ginzburg-Landau-level description of thermodynamics, dynamics, and noise of a disordered noncentrosymmetric superconductor, taking the two-dimensional Rashba model with an in-plane Zeeman field as a representative example, within the Keldysh nonlinear sigma model.

On the thermodynamic side, the theory contains the Lifshitz invariants of the free energy with explicit disorder dependence: the linear invariant $\bm\Lambda$, the cubic gradient invariant $\Xi$, and the momentum-odd quartic vertex $\bm{\mathfrak b}$, all directed along $\bm n=[\bm h\times\bm\alpha]$. The couplings of the quadratic sector are determined by a single closed-form kernel, Eq.~\eqref{eq:Kkernel}, whose argument is the ratio of the Dyakonov-Perel rate to temperature; the odd quartic vertex is taken from the diagrammatic calculation of Ref.~\cite{Hasan:2025} in the weak-relaxation regime. The kernel interpolates between the weak relaxation regime, where every invariant is suppressed as $(\alpha_R p_F\tau)^2$, and the relaxation-dominated regime, where the helical wave vector saturates at the universal value $q_0=4\alpha_R h/v_F^2$. This crossover, Eq.~\eqref{eq:gfun} and Fig.~\ref{fig:crossover}, reconciles previously conflicting results for the magnetoelectric response of dirty Rashba superconductors.

On the dynamical side unique to the Keldysh formulation, we found that the same microscopic physics places a nonreciprocal term into the kinetic coefficient of the TDGL equation and, through the fluctuation-dissipation theorem (which holds in the presence of all magnetoelectric couplings), into the Langevin noise, Eq.~\eqref{eq:noisefinal}. The structure of this kinetic Lifshitz invariant is fixed by Onsager reciprocity: friction and noise change together, so the equal-time fluctuation spectrum remains Gibbsian, nonreciprocal only through the thermodynamic invariants, and the Bloch theorem (no current in equilibrium) is respected, while the relaxation dynamics and the nonlinear response acquire the additional, kinetic nonreciprocity. In the fluctuation-induced magnetochiral anisotropy the kinetic invariant contributes on equal footing with the cubic thermodynamic invariant, Eq.~\eqref{eq:MCAresult}, once both are referred to the helical minimum; the amplitude is governed by the irreducibility factor $R(\Gamma_t/4\pi T)$ of Eq.~\eqref{eq:Rfactor}, equal to $0.41$ at weak relaxation and vanishing beyond the Dyakonov-Perel crossover, and the nonreciprocal resistance grows as $\epsilon^{-2}$ on approach to $T_c$, Eq.~\eqref{eq:MCAgeneral}, becoming independent of temperature at fixed current where the fluctuation conductivity dominates, Eq.~\eqref{eq:MCAj}. These results have a common origin: in the relaxation-dominated regime the odd part of the pair kernel becomes proportional to the derivative of its even part, Eq.~\eqref{eq:removability}, so that all first-order nonreciprocity of the Gaussian sector reduces to the helical shift. The helical modulation itself remains observable, but transport coefficients and noise, which involve an integral over all pair momenta, are not affected by it; a nonreciprocal response of this kind requires either weak spin relaxation or the quartic vertex. In the diode effect below $T_c$ the cubic invariant competes with the momentum-odd quartic vertex and with the even higher-gradient terms promoted to odd ones by the helical shift; the net result, Eq.~\eqref{eq:eta}, is a reversal of the diode polarity relative to truncated single- or two-term estimates, in agreement with the diffusive limit of the diagrammatic theory \cite{Hasan:2024,Hasan:2025}.

Several directions follow naturally. Out of equilibrium the FDT no longer constrains nonreciprocal friction to noise, and the Keldysh block acquires independent nonreciprocal content. Driven or current-biased noncentrosymmetric films are thus a natural platform for noise rectification and fluctuation ratchets. Below $T_c$ the interplay of the helical ground state with phase slips, vortices, and the Maki-Thompson channel of nonreciprocal transport remains open. The same machinery extends to other parity-breaking structures (Dresselhaus coupling, Ising spin-orbit coupling of transition-metal dichalcogenides), to quasi-2D films and wires, and to higher cumulants of the current noise \cite{Bagrets:2014}, where the kinetic Lifshitz invariant may appear in the third cumulant already in equilibrium.

%==============================================================================================================================================================================================
%==============================================================================================================================================================================================
%==============================================================================================================================================================================================
%==============================================================================================================================================================================================
%==============================================================================================================================================================================================
\section*{Acknowledgements}

We are grateful to Ilya Tokatly for several useful discussions and for sharing his notes on the derivation of the Usadel equation.
The work of T.~L., J.~T.~M. and D.~S. was supported in part by the NSF Quantum Leap Challenge Institute for Hybrid Quantum Architectures and Networks Grant No. OMA-2016136. 
The work of J. T. M. was supported in addition by the Coordenação de Aperfeiçoamento de Pessoal de Nível Superior - Brasil (CAPES) - Finance Code 001. 
The work of A.~L. was supported by NSF Grant No. DMR-2452658 and H.~I. Romnes Faculty Fellowship provided by the University of Wisconsin--Madison Office of the Vice Chancellor for Research and Graduate Education with funding from the Wisconsin Alumni Research Foundation. This work employed the large language model Claude (Anthropic) in the preparation of the manuscript, including in particular, graphics, and symbolic verification of analytical results. The authors conceived the project and carried out the
research.

%==============================================================================================================================================================================================
%==============================================================================================================================================================================================
%==============================================================================================================================================================================================
%==============================================================================================================================================================================================
%==============================================================================================================================================================================================
\appendix

\section{Cooperon expansion and the Gaussian action}
\label{app:gaussian}

In this appendix we derive the kernel \eqref{eq:Amatrix} and the sources \eqref{eq:sources}. All matrix traces below were also checked by a symbolic computation in the $8\times8$ Keldysh$\otimes$Nambu$\otimes$spin space, with the full distribution-function content of $U$ and $\Lambda$ retained and the quadratic forms symmetrized over $(\bm q,\e,\e')\to(-\bm q,\e',\e)$.

(i) \emph{Expansion of $Q$}.
Since $\{w,\hat\sigma_3\hat\tau_3\}=0$, the parametrization \eqref{eq:param} gives exactly $Q=U\hat\sigma_3\hat\tau_3\,e^wU^{-1}$, so that
\begin{equation}
Q^{(1)}=U\hat\sigma_3\hat\tau_3\,w\,U^{-1},
\qquad
Q^{(2)}=\tfrac12\,U\hat\sigma_3\hat\tau_3\,w^2\,U^{-1}.
\end{equation}

(ii) \emph{Gradient term}. 
With $[\mathcal A_k,\Lambda]=0$ (the saddle point is a spin scalar), the covariant gradient of $Q$ starts at first order, $\tilde\nabla_kQ=\nabla_kQ^{(1)}-i[\mathcal A_k,Q^{(1)}]+\dots$, and
\begin{equation}
D\Tr(\tilde\nabla Q)^2
=D\Tr\big[(\nabla Q^{(1)})^2\big]
-2iD\Tr\big[\nabla Q^{(1)}[\mathcal A,Q^{(1)}]\big]
-D\Tr\big[[\mathcal A,Q^{(1)}][\mathcal A,Q^{(1)}]\big].
\end{equation}
The first term yields the diagonal $Dq^2$; the second the precession entries $4iDm\alpha_R q_{y,z}$ between $c_1$ and $c_{2,3}$ (the singlet is protected, $[\mathcal A_k,\hat s_0]=0$); the third the DP matrix $\mathrm{diag}(0,2\Gamma_t,\Gamma_t,\Gamma_t)$, $\Gamma_t=4Dm^2\alpha_R^2$, via $[\hat s_a,[\hat s_a,\hat s_b]]=4\hat s_b(1-\delta_{ab})$ summed over the spin directions $a\in\{2,3\}$ present in $\mathcal A_{y,z}$.

(iii) \emph{Time derivative}.
The term $4i\Tr[i\hat\tau_3\partial_tQ]$, evaluated with $Q^{(2)}$, gives $-i(\e+\e')$ in the $c$ sector and $+i(\e+\e')$ in the $\bar c$ sector: the retarded and advanced cooperons. (The opposite signs are essential; with equal signs the theory would not produce the correct causality structure of $L_{R,A}$.)

(iv) \emph{Zeeman term}.
For $4ih\Tr[\hat\tau_3\hat s_3Q^{(2)}]=2ih\Tr[\hat\sigma_3\hat s_3w^2]$ one uses $\Tr_s[\hat s_3\hat s_\mu\hat s_\nu]\ne0$ for $(\mu\nu)\in\{(03),(30),(12),(21)\}$. The $(12)$ channel cancels upon symmetrization of the quadratic form, while the $(03)$ channel survives with equal coefficients for the two orderings: the coupling is symmetric, $\mathcal C_{03}=\mathcal C_{30}=-2ih$ (retarded sector). The physical check via the pole structure, Eq.~\eqref{eq:zeemanpoles}, and its integrated form
\begin{equation}
\ln\frac{T_c}{T}
-\frac12\Big[\psi\Big(\frac12+x+\frac{ih}{2\pi T}\Big)+\psi\Big(\frac12+x-\frac{ih}{2\pi T}\Big)\Big]
+\psi\Big(\frac12\Big)
\simeq\ln\frac{T_c}{T}-\frac{7\zeta(3)h^2}{4\pi^2T^2}+\ldots
\label{eq:hsquare}
\end{equation}
(valid at $\Gamma_t\to0$) fix both the sign and the symmetry.

(v) \emph{Field-strength term}.
At Gaussian order the last term of \eqref{eq:S} requires $Q\to\Lambda$ in the undifferentiated slot and the mixed commutator-gradient contribution in the two covariant derivatives:
\begin{equation}
\varkappa_g\Tr\Big[F_{yz}\,\Lambda\big(\tilde\nabla_yQ^{(1)}\tilde\nabla_zQ^{(1)}
-\tilde\nabla_zQ^{(1)}\tilde\nabla_yQ^{(1)}\big)\Big]
\;\Rightarrow\;
\begin{cases}
\mathcal C^{(F)}_{03}=\mathcal C^{(F)}_{30}=+4i\kappa_0q_y,\\[2pt]
\mathcal C^{(F)}_{02}=\mathcal C^{(F)}_{20}=-4i\kappa_0q_z ,
\end{cases}
\end{equation}
with $\kappa_0=m^3\alpha_R^3\varkappa_g$. A word of caution: when symmetrizing the quadratic form under $\bm q\to-\bm q$, the explicit momentum factor in these couplings also changes sign, so the couplings remain symmetric in the spin indices; treating them as antisymmetric flips the sign of the $h\kappa_0$ cross term in \eqref{eq:Ainv} and, together with the pole structure of the Zeeman channel, is excluded by the physical checks above.

(vi) \emph{Sector structure}.
Both the Zeeman and field-strength couplings reverse sign between the $c$ and $\bar c$ sectors, so their product does not; hence $L_A^{-1}=[L_R^{-1}]^*$ holds with all magnetoelectric terms included. The sources \eqref{eq:sources} follow from the first-order expansion of $4i\Tr[\Delta Q]$ and carry the distribution functions $f_\e$ (retarded) and $f_{\e'}$ (advanced) exactly as displayed; these assignments generate, upon Gaussian integration, the $f_{\e\mp\omega/2}$ structures of Eq.~\eqref{eq:Ldefs}.

%==============================================================================================================================================================================================
%==============================================================================================================================================================================================
%==============================================================================================================================================================================================
%==============================================================================================================================================================================================
%==============================================================================================================================================================================================
\section{Master integrals}
\label{app:integrals}

With $x_k=\pi(k+\tfrac12)$ the Matsubara decomposition $\tanh x=\sum_{k\ge0}2x/(x^2+x_k^2)$ and closure of the contour around the poles $x=ix_k$ give
\begin{equation}
\int_{-\infty}^{\infty}\!dx\,\frac{\tanh x}{(z-ix)^n}
=2\pi i\sum_{k\ge0}\frac{1}{(z+x_k)^n}
=\frac{2\pi i\,(-1)^n}{\pi^n(n-1)!}\,
\psi^{(n-1)}\Big(\frac12+\frac{z}{\pi}\Big),
\label{eq:tanhint}
\end{equation}
where the second equality uses $\psi^{(m)}(y)=(-1)^{m+1}m!\sum_{k\ge0}(k+y)^{-m-1}$. Equation \eqref{eq:tanhint}, rescaled by $\e=2Tx$, produces the list \eqref{eq:Jn}; the values needed in the text are
\begin{equation}
\psi'(\tfrac12)=\frac{\pi^2}{2},\qquad
\psi^{(2)}(\tfrac12)=-14\zeta(3),\qquad
\psi^{(3)}(\tfrac12)=\pi^4,\qquad
\psi^{(4)}(\tfrac12)=-744\zeta(5),
\end{equation}
following from $\psi^{(n)}(\tfrac12)=(-1)^{n+1}n!(2^{n+1}-1)\zeta(n+1)$. The crossover kernel \eqref{eq:Kkernel} follows from partial fractions as described in the text; its small-$g$ limit uses $\psi(\tfrac12+x+g)-\psi(\tfrac12+x)=g\psi'+\tfrac{g^2}{2}\psi''+O(g^3)$, whose linear term cancels inside the bracket of \eqref{eq:Kkernel} leaving $-\tfrac{g}{2}\psi''$, i.e., $\mathcal K\to-\psi''(\tfrac12+x)/2(4\pi T)^2$; the large-$g$ limit follows from $[\psi(\tfrac12+x+g)-\psi(\tfrac12+x)]/g\sim(\ln g)/g$. 

%==============================================================================================================================================================================================
%==============================================================================================================================================================================================
%==============================================================================================================================================================================================
%==============================================================================================================================================================================================
%==============================================================================================================================================================================================
\section{Quartic vertex}
\label{app:quartic}

(i) \emph{Uniform coefficient.} Expanding $\Tr[\Delta Q]$ to third order in $w$ and substituting the Gaussian saddle point of the cooperons, $c_\mu=-2i\mathcal C^{-1}_{\mu\nu}(\Delta^{c}_\nu+f_\e\Delta^q_\nu)$ and its conjugate, generates the quartic contribution to the action with the kernel $-i\int d\e\,f_{\e-\omega/2}\,[\mathcal C^{-1}_{00}(\bm q,\e)]^3$ in the singlet channel. At $\bm q=h=0$, using $J_3$ of \eqref{eq:Jn}, this yields the standard $b_0=7\zeta(3)/8\pi^2T^2$.

(ii) \emph{Gradient terms.} It is tempting to obtain the momentum dependence of $b$ by evaluating the same kernel at the shifted cooperon pole (``all pair momenta near a common $\bm q$''). This shortcut is incorrect for the gradient corrections: the full vertex is a convolution over three independent pair momenta, and the pieces it misses (in diagrammatic language, impurity ladders connecting nonadjacent sections of the quartic loop, and the triplet-exchange insertions) contribute at the same order in $Dq^2/T$. The even sector provides a clean demonstration, because there the exact answer is available from the nonlinear saddle point of the sigma model, i.e., the Usadel equation. For the plane-wave state $\Delta=|\Delta|e^{i\bm q\cdot\bm r}$ the Usadel problem is that of a uniform pair breaker $s=Dq^2/2$: with the standard angular parametrization,
\begin{align}
&\omega_n\sin\theta_n+s\,\sin\theta_n\cos\theta_n=|\Delta|\cos\theta_n ,
\nonumber\\
&\frac{F}{\nu}=2\pi T\!\!\sum_{\omega_n>0}\!\Big[2\omega_n(1-\cos\theta_n)+s\sin^2\theta_n-2|\Delta|\sin\theta_n\Big]+\frac{|\Delta|^2}{\lambda} ,
\end{align}
where $\lambda$ is the pairing constant. Expanding with $\theta_n=|\Delta|/(\omega_n+s)+O(|\Delta|^3)$, the quadratic term reproduces the Abrikosov-Gor'kov kernel of Eq.~\eqref{eq:LRweak} exactly, $1/\lambda-2\pi T\sum_{\omega_n>0}(\omega_n+s)^{-1}\to\epsilon+\psi(\tfrac12+x)-\psi(\tfrac12)$ with $x=s/2\pi T=Dq^2/4\pi T$, including the $q^4$ coefficient $\alpha_4=7\zeta(3)(D/4\pi T)^2$ used in Sec.~\ref{sec:SDE}, while the quartic term collects to
\begin{equation}
\frac{F_4}{\nu}=|\Delta|^4\,2\pi T\sum_{\omega_n>0}\frac{\omega_n}{4\,(\omega_n+s)^4}
\;\;\Longrightarrow\;\;
b(q)=b_0-\frac{\pi D}{48\,T^3}\,q^2+O(q^4),
\end{equation}
i.e., $\mathfrak b_2=\pi^3D/42\zeta(3)T$ of Eq.~\eqref{eq:bq}. Note the per-frequency weight $\omega_n/(\omega_n+s)^4$: the single-mode substitution into the uniform kernel would give $\sum(\omega_n+s)^{-3}$, whose slope is smaller by the factor $3/4$. The two expressions agree at $q=0$ (Anderson theorem) but differ in every gradient correction.

(iii) \emph{Odd part.} At first order in the nonreciprocal coupling the vertex acquires, besides the shifted-pole contribution, triplet-assisted corrections of the same order (one Zeeman conversion $h$ and one field-strength vertex $\kappa_0 q$ inserted anywhere on the quartic loop). Rather than organizing this expansion within the sigma model, we take the coefficient from the diagrammatic evaluation of the complete set of quartic diagrams at weak spin-orbit coupling and arbitrary disorder, Ref.~\cite{Hasan:2025}: in our conventions [$\beta(q)=\tfrac{\nu}{2}b(q)$] the diffusive limit of Eq.~(C4c) of that work gives the odd term of Eq.~\eqref{eq:bq}, $b_0\mathfrak b=31\zeta(5)(\alpha_R p_F\tau)^2\alpha_R h/2\pi^4T^4$, larger than the single-mode estimate [$93\zeta(5)/14\zeta(3)\pi^2$ in place of $124\zeta(5)/7\zeta(3)\pi^2$] by the factor $8/3$. As a check, the full coefficient set $\{\Lambda,\Xi,\alpha_4,b_0,\mathfrak b,\mathfrak b_2\}$ substituted into the general diode formula of Ref.~\cite{Hasan:2024} reproduces the diffusive-limit diode efficiency of Ref.~\cite{Hasan:2025} (Sec.~\ref{sec:SDE}); the couplings $\Lambda$, $\Xi$, and $\alpha_4$ obtained here from the sigma model agree term by term with the diffusive limits of Eqs.~(C4a), (C4b), and (C3b) of Ref.~\cite{Hasan:2025}.

%==============================================================================================================================================================================================
%==============================================================================================================================================================================================
%==============================================================================================================================================================================================
%==============================================================================================================================================================================================
%==============================================================================================================================================================================================
\section{Nonlinear fluctuation transport}
\label{app:MCA}

Here we outline the solution of the kinetic equation \eqref{eq:kinetic} and the rectification integrals behind Eq.~\eqref{eq:MCAresult}. We first present the expansion with the truncated inputs $a=\epsilon+\xi^2_{GL}q^2-\Lambda q_n+\Xi q_nq^2$ and 
$\Gamma_\rho=1+\bm\rho\cdot\bm q$, which gives the weights of $\bm\rho$ and $\Xi$ and the cancellation of the $\Lambda$ terms, and then add the even gradient corrections $-\alpha_4q^4$ and $\rho_2q^2$, through which $\Lambda$ contributes at the same order. Iterating in the field, $n=n^{(0)}+n^{(1)}+n^{(2)}$ with $n^{(0)}=T/\nu a$ and $n^{(k)}=-\lambda_{\bm q}^{-1}\,e\bm E\cdot\partial_{\bm q}n^{(k-1)}$, and writing $1/\lambda_{\bm q}=(\pi/8T)(1+\bm\rho\cdot\bm q)/a$, the current density along $\hat{\bm n}$ becomes ($\beta\equiv\pi eE/8$)
\begin{equation}
j^{(2)}=-\frac{2e\beta^2}{T}\int\!\frac{d^2q}{(2\pi)^2}\,
\Gamma_\rho\,\frac{a'_n}{a}\,
\partial_n\Big[\Gamma_\rho\frac{a'_n}{a^3}\Big],
\qquad \Gamma_\rho=1+\bm\rho\cdot\bm q .
\end{equation}
Expanding to first order in the odd couplings, with $a=A+\delta a$, $A=\epsilon+\xi^2_{GL}q^2$, $\delta a=-\Lambda q_n+\Xi\,q_nq^2$, and using the angular averages $\langle q_n^2\rangle=q^2/2$, $\langle q_n^4\rangle=3q^4/8$, $\langle q_n^2q_t^2\rangle=q^4/8$ together with the radial integrals
\begin{equation}
\int\!\frac{d^2q}{(2\pi)^2}\,\frac{(q^2)^s}{A^m}
=\frac{1}{4\pi\,\xi_{GL}^{2s+2}\,\epsilon^{\,m-s-1}}\,
\frac{s!\,(m-s-2)!}{(m-1)!} ,
\end{equation}
one finds the following list of first-order contributions (in units of $1/\pi\epsilon^2$ for $\rho$ and $\Xi/\xi^2_{GL}$, and of $1/\pi\epsilon^3$ for $\Lambda$):
\begin{center}
\begin{tabular}{lccc}
\toprule
source & $\rho$ & $\Xi/\xi^2_{GL}$ & $\Lambda$ \\
\midrule
$\bm\rho\cdot\bm q$ term of $\Gamma_\rho$ (two placements) & $\tfrac1{12}+\tfrac16$ & --- & --- \\
$\delta a$ in vertices and denominators & $-\tfrac38$ & $\tfrac16+\tfrac14-\tfrac16-\tfrac{15}{16}+\tfrac9{16}$ & $-\tfrac16+\tfrac16+\tfrac38-\tfrac38$ \\
\midrule
total & $-\tfrac18$ & $-\tfrac18$ & $0$ \\
\bottomrule
\end{tabular}
\end{center}
The vanishing of the $\Lambda$ column is the momentum-space counterpart of the removability of the linear invariant; the equality of the $\rho$ and $\Xi/\xi^2_{GL}$ totals is specific to two dimensions.

The vanishing of the $\Lambda$ column does not yet mean that $\Lambda$ drops out of the result. The $\Lambda$-linear terms are individually of order $1/\epsilon^3$, one power more singular than the $\rho$ and $\Xi$ terms, so that with a single insertion of an even gradient correction, $-\alpha_4q^4$ in $a$ or $\rho_2q^2$ in $\Gamma_\rho$, the $\Lambda$-dependent cross terms are of order $1/\epsilon^2$, the same as the $\rho$ and $\Xi$ terms; all further insertions are suppressed by additional powers of $\epsilon$. Repeating the expansion with $A=\epsilon+\xi^2_{GL}q^2-\alpha_4q^4$ and $1/\lambda_{\bm q}=(\pi/8T)(1+\bm\rho\cdot\bm q+\rho_2q^2)/a$, the two additional columns have the totals (in the same units, $1/\pi\epsilon^2$)
\begin{equation}
\Lambda\alpha_4/\xi^4_{GL}:\;+\tfrac14,
\qquad
\Lambda \rho_2/\xi^2_{GL}:\;-\tfrac18,
\end{equation}
while the $\rho$ and $\Xi$ totals are unchanged at this order. Hence $\rho+\Xi/\xi^2_{GL}\to\rho+\Xi/\xi^2_{GL}-2\Lambda\alpha_4/\xi^4_{GL}+\Lambda \rho_2/\xi^2_{GL}=\rho_{\rm eff}+\Xi_{\rm eff}/\xi^2_{GL}$, which is Eq.~\eqref{eq:effcouplings}; with $\rho_2=-2\alpha_4/\xi^2_{GL}$ the two cross terms combine to $-4\Lambda\alpha_4/\xi^4_{GL}$. Equivalently, performing the shift $\bm q\to\bm q+q_0\hat{\bm n}$ before expanding removes $\Lambda$ altogether, and the table above applies directly to the effective couplings. Restoring prefactors,
\begin{equation}
j^{(2)}=\frac{2e\beta^2}{T}\cdot\frac{\rho_{\rm eff}+\Xi_{\rm eff}/\xi^2_{GL}}{8\pi\,\epsilon^2}
=\frac{\pi e^3E^2}{256\,T}\,\frac{\rho_{\rm eff}+\Xi_{\rm eff}/\xi^2_{GL}}{\epsilon^2},
\end{equation}
which is Eq.~\eqref{eq:MCAresult}. At first order in $E$ the same machinery gives $j^{(1)}=e^2E/16\epsilon$, the AL result, fixing the sign conventions and normalizations. In a quasi-two-dimensional film of thickness $d$ all results acquire the usual factor $1/d$, and the relative weights of $\rho$ and $\Xi$ are modified by order-one factors.

Finally, we give the details of two statements made in Sec.~\ref{sec:tdgleq}. 

(i) \emph{The current.} With the vector potential on the two branches of the contour written as $\bm A_\pm=\bm A^c\pm\bm A^q$, the current is $\bm j=\tfrac12\langle\delta S_{GL}/\delta\bm A^q\rangle$. The quantum component enters the quadratic action \eqref{eq:SGL} through the Keldysh matrix structure, so that the vertex $\partial L_R^{-1}/\partial\bm A=-2e\,\partial L_R^{-1}/\partial\bm q$ of the retarded block and the corresponding vertex of the advanced block are both transferred to the classical--classical component, $\bm j=\tfrac{\nu}{2}\langle\Delta^{*c}\,[\partial_{\bm A}L_R^{-1}+\partial_{\bm A}L_A^{-1}]\,\Delta^{c}\rangle$ (the quantum--quantum correlator vanishes). The static part of the kernel, $-a(\bm q)$, is common to the two blocks and gives the thermodynamic vertex $2e\nu\,\partial_{q_n}a$ of Eq.~\eqref{eq:kinetic}, whereas the dissipative part, $\pm i\omega\gamma(\bm q)$, has opposite signs in $L_R^{-1}$ and $L_A^{-1}$ and drops out. The momentum dependence of the friction therefore does not generate a current vertex of its own, and $j=2e\nu\int_{\bm q}a'_n n_{\bm q}$ holds with the nonreciprocal friction included. 

(ii) \emph{The ordering.} In the gauge $\bm A=-\bm Et$ the mode of canonical momentum $\bm q$ has kinetic momentum $\bm k(t)=\bm q+2e\bm Et$, and the mixed term of Eq.~\eqref{eq:TDGL} can be written as $\Gamma_\rho(\bm k)\partial_t\Delta$ [as written in \eqref{eq:TDGL}], as $\partial_t[\Gamma_\rho(\bm k)\Delta]$, or as their average; the three forms differ by $c\,\partial_t\Gamma_\rho\,\Delta=2ce\bm E\cdot\partial_{\bm k}\Gamma_\rho\,\Delta$ with $c=0,1,\tfrac12$, which adds the term $-2c\,(2e\bm E\cdot\partial_{\bm k}\Gamma_\rho/\Gamma_\rho)\,n_{\bm q}$ to the right-hand side of the kinetic equation \eqref{eq:kinetic} and is of the same order as the terms retained there. Repeating the expansion of this appendix with that term, the $\rho$ column of the table changes from $-\tfrac18$ to $-\tfrac18(1-4c)$, i.e., to $+\tfrac18$ for the symmetrized form and $+\tfrac38$ for $\partial_t[\Gamma_\rho\Delta]$, while the $\Xi$ column is unaffected; since $\rho=\Xi/\xi^2_{GL}$, the symmetrized form would thus give no magnetochiral anisotropy. The ordering is fixed by local detailed balance: only for $c=0$ does the stationary distribution of a mode at fixed $\bm k$ coincide with the Gibbs distribution $\propto\exp[-\nu a(\bm k)|\Delta|^2/T]$ of its instantaneous free energy, while for $c\neq0$ the extra term shifts the effective $a$ by an amount $2ce\bm E\cdot\partial_{\bm k}\Gamma_\rho\times\pi/8T$ not accompanied by any change of the noise. Local detailed balance with respect to the instantaneous free energy is the condition under which the stochastic dynamics obeys the fluctuation theorem, as the underlying microscopic theory does. We have checked this directly by computing the generating function $\mathcal F(\chi;E)$ of the charge transported by the fluctuation current in the driven ensemble: for $c=0$ it satisfies the global detailed-balance property $\mathcal F(-E/T;E)=0$ of Ref.~\cite{Forster:2008} and the field-reversing symmetry $\mathcal F(\chi;E;\bm h)=\mathcal F(-\chi-E/T;E;-\bm h)$ of Ref.~\cite{SaitoUtsumi:2008}, together with the resulting relations between the rectified current, the field-odd noise, and the equilibrium third cumulant, whereas for $c=\tfrac12$ and $c=1$ these properties fail, and the third-cumulant relation fails in particular because the equilibrium third cumulant does not depend on the ordering.

\bibliographystyle{elsarticle-num}
\bibliography{biblio-Lifshitz}

\begin{thebibliography}{10}
\expandafter\ifx\csname url\endcsname\relax
  \def\url#1{\texttt{#1}}\fi
\expandafter\ifx\csname urlprefix\endcsname\relax\def\urlprefix{URL }\fi
\expandafter\ifx\csname href\endcsname\relax
  \def\href#1#2{#2} \def\path#1{#1}\fi

\bibitem{Landau:1937}
L.~D. Landau, On the theory of phase transitions, Zh. Eksp. Teor. Fiz. 7 (1937)
  19, [Phys. Z. Sowjetunion {\bf 11}, 26 (1937)].

\bibitem{Lifshitz:1941}
E.~M. Lifshitz, On the theory of phase transitions of the second order i \& ii,
  Zh. Eksp. Teor. Fiz. 11 (1941) 255, 269.

\bibitem{Dzyaloshinsky:1958}
I.~Dzyaloshinsky, A thermodynamic theory of ``weak'' ferromagnetism of
  antiferromagnetics, J. Phys. Chem. Solids 4 (1958) 241.

\bibitem{Moriya:1960}
T.~Moriya, Anisotropic superexchange interaction and weak ferromagnetism, Phys.
  Rev. 120 (1960) 91.

\bibitem{Bogdanov:1989}
A.~N. Bogdanov, D.~A. Yablonskii, Thermodynamically stable ``vortices'' in
  magnetically ordered crystals. the mixed state of magnets, Zh. Eksp. Teor.
  Fiz. 95 (1989) 178, [Sov. Phys. JETP {\bf 68}, 101 (1989)].

\bibitem{Muhlbauer:2009}
S.~M{\"u}hlbauer, B.~Binz, F.~Jonietz, C.~Pfleiderer, A.~Rosch, A.~Neubauer,
  R.~Georgii, P.~B{\"o}ni, Skyrmion lattice in a chiral magnet, Science 323
  (2009) 915.

\bibitem{deGennes:1993}
P.~G. de~Gennes, J.~Prost, The Physics of Liquid Crystals, Oxford University
  Press, Oxford, 1993.

\bibitem{Zubko:2013}
P.~Zubko, G.~Catalan, A.~K. Tagantsev, Flexoelectric effect in solids, Annu.
  Rev. Mater. Res. 43 (2013) 387.

\bibitem{BauerSigrist:2012}
E.~Bauer, M.~Sigrist (Eds.), Non-Centrosymmetric Superconductors: Introduction
  and Overview, Lecture Notes in Physics Vol. 847, Springer, Berlin, 2012.

\bibitem{Smidman:2017}
M.~Smidman, M.~B. Salamon, H.~Q. Yuan, D.~F. Agterberg, Superconductivity and
  spin-orbit coupling in non-centrosymmetric materials: a review, Rep. Prog.
  Phys. 80 (2017) 036501.

\bibitem{MineevSamokhin:1994}
V.~P. Mineev, K.~V. Samokhin, Helical phases in superconductors, Zh. Eksp.
  Teor. Fiz. 105 (1994) 747, [JETP {\bf 78}, 401 (1994)].

\bibitem{Agterberg:2003}
D.~F. Agterberg, Novel magnetic field effects in unconventional
  superconductors, Physica C 387 (2003) 13.

\bibitem{Barzykin:2002}
V.~Barzykin, L.~P. Gor'kov, Inhomogeneous stripe phase revisited for surface
  superconductivity, Phys. Rev. Lett. 89 (2002) 227002.

\bibitem{Kaur:2005}
R.~P. Kaur, D.~F. Agterberg, M.~Sigrist, Helical vortex phase in the
  noncentrosymmetric {CePt$_3$Si}, Phys. Rev. Lett. 94 (2005) 137002.

\bibitem{DimitrovaFeigelman:2007}
O.~Dimitrova, M.~V. Feigel'man, Theory of a two-dimensional superconductor with
  broken inversion symmetry, Phys. Rev. B 76 (2007) 014522.

\bibitem{Samokhin:2004}
K.~V. Samokhin, Magnetic properties of superconductors with strong spin-orbit
  coupling, Phys. Rev. B 70 (2004) 104521.

\bibitem{MineevSamokhin:2008}
V.~P. Mineev, K.~V. Samokhin, Effects of impurities on superconductivity in
  noncentrosymmetric compounds, Phys. Rev. B 78 (2008) 144503.

\bibitem{Yip:2002}
S.~K. Yip, Two-dimensional superconductivity with strong spin-orbit
  interaction, Phys. Rev. B 65 (2002) 144508.

\bibitem{Buzdin:2008}
A.~Buzdin, Direct coupling between magnetism and superconducting current in the
  {Josephson} $\varphi_0$ junction, Phys. Rev. Lett. 101 (2008) 107005.

\bibitem{Konschelle:2015}
F.~Konschelle, I.~V. Tokatly, F.~S. Bergeret, Theory of the spin-galvanic
  effect and the anomalous phase shift $\varphi_0$ in superconductors and
  {Josephson} junctions with intrinsic spin-orbit coupling, Phys. Rev. B 92
  (2015) 125443.

\bibitem{Assouline:2019}
A.~Assouline, C.~Feuillet-Palma, N.~Bergeal, T.~Zhang, A.~Mottaghizadeh,
  A.~Zimmers, E.~Lhuillier, M.~Eddrie, P.~Atkinson, M.~Aprili, J.~Lesueur,
  Spin-orbit induced phase-shift in {Bi$_2$Se$_3$} {Josephson} junctions, Nat.
  Commun. 10 (2019) 126.

\bibitem{Edelstein:1989}
V.~M. Edelstein, Characteristics of the {Cooper} pairing in two-dimensional
  noncentrosymmetric electron systems, Zh. Eksp. Teor. Fiz. 95 (1989) 2151,
  [Sov. Phys. JETP {\bf 68}, 1244 (1989)].

\bibitem{Edelstein:1995}
V.~M. Edelstein, Magnetoelectric effect in polar superconductors, Phys. Rev.
  Lett. 75 (1995) 2004.

\bibitem{GorkovRashba:2001}
L.~P. Gor'kov, E.~I. Rashba, Superconducting {2D} system with lifted spin
  degeneracy: mixed singlet-triplet state, Phys. Rev. Lett. 87 (2001) 037004.

\bibitem{Ando:2020}
F.~Ando, Y.~Miyasaka, T.~Li, J.~Ishizuka, T.~Arakawa, Y.~Shiota, T.~Moriyama,
  Y.~Yanase, T.~Ono, Observation of superconducting diode effect, Nature 584
  (2020) 373.

\bibitem{Baumgartner:2022}
C.~Baumgartner, L.~Fuchs, A.~Cost{\'a}, S.~Reinhardt, S.~Gronin, G.~C. Gardner,
  T.~Lindemann, M.~J. Manfra, P.~E. Faria~Junior, D.~Kochan, J.~Fabian,
  N.~Paradiso, C.~Strunk, Supercurrent rectification and magnetochiral effects
  in symmetric {Josephson} junctions, Nat. Nanotechnol. 17 (2022) 39.

\bibitem{Nadeem:2023}
M.~Nadeem, M.~S. Fuhrer, X.~Wang, The superconducting diode effect, Nat. Rev.
  Phys. 5 (2023) 558.

\bibitem{Shaffer:2025}
D.~Shaffer, A.~Levchenko, Theories of superconducting diode effects (2025).
\newblock \href {http://arxiv.org/abs/2510.25864} {\path{arXiv:2510.25864}}.

\bibitem{Daido:2022}
A.~Daido, Y.~Ikeda, Y.~Yanase, Intrinsic superconducting diode effect, Phys.
  Rev. Lett. 128 (2022) 037001.

\bibitem{YuanFu:2022}
N.~F.~Q. Yuan, L.~Fu, Supercurrent diode effect and finite-momentum
  superconductors, Proc. Natl. Acad. Sci. USA 119 (2022) e2119548119.

\bibitem{HeTanakaLaw:2022}
J.~J. He, Y.~Tanaka, K.~T. Law, A phenomenological theory of superconductor
  diodes, New J. Phys. 24 (2022) 053014.

\bibitem{IlicBergeret:2022}
S.~Ili{\'c}, F.~S. Bergeret, Theory of the supercurrent diode effect in
  {Rashba} superconductors with arbitrary disorder, Phys. Rev. Lett. 128 (2022)
  177001.

\bibitem{Hasan:2025}
J.~Hasan, D.~Shaffer, M.~Khodas, A.~Levchenko, Superconducting diode efficiency
  from singlet-triplet mixing in disordered systems, Phys. Rev. B 111 (2025)
  174514.

\bibitem{Hasan:2024}
J.~Hasan, D.~Shaffer, M.~Khodas, A.~Levchenko, Supercurrent diode effect in
  helical superconductors, Phys. Rev. B 110 (2024) 024508.

\bibitem{Rikken:2001}
G.~L. J.~A. Rikken, J.~F{\"o}lling, P.~Wyder, Electrical magnetochiral
  anisotropy, Phys. Rev. Lett. 87 (2001) 236602.

\bibitem{Rikken:2005}
G.~L. J.~A. Rikken, P.~Wyder, Magnetoelectric anisotropy in diffusive
  transport, Phys. Rev. Lett. 94 (2005) 016601.

\bibitem{TokuraNagaosa:2018}
Y.~Tokura, N.~Nagaosa, Nonreciprocal responses from non-centrosymmetric quantum
  materials, Nat. Commun. 9 (2018) 3740.

\bibitem{Wakatsuki:2017}
R.~Wakatsuki, Y.~Saito, S.~Hoshino, Y.~M. Itahashi, T.~Ideue, M.~Ezawa,
  Y.~Iwasa, N.~Nagaosa, Nonreciprocal charge transport in noncentrosymmetric
  superconductors, Sci. Adv. 3 (2017) e1602390.

\bibitem{Itahashi:2020}
Y.~M. Itahashi, T.~Ideue, Y.~Saito, S.~Shimizu, T.~Ouchi, T.~Nojima, Y.~Iwasa,
  Nonreciprocal transport in gate-induced polar superconductor {SrTiO$_3$},
  Sci. Adv. 6 (2020) eaay9120.

\bibitem{WakatsukiNagaosa:2018}
R.~Wakatsuki, N.~Nagaosa, Nonreciprocal current in noncentrosymmetric {Rashba}
  superconductors, Phys. Rev. Lett. 121 (2018) 026601.

\bibitem{Hoshino:2018}
S.~Hoshino, R.~Wakatsuki, K.~Hamamoto, N.~Nagaosa, Nonreciprocal charge
  transport in two-dimensional noncentrosymmetric superconductors, Phys. Rev. B
  98 (2018) 054510.

\bibitem{DeMiranda:2026}
J.~T. de~Miranda, M.~Khodas, A.~Levchenko, Electrical magnetochiral anisotropy
  in {Rashba} superconductors, Low Temperature Physics 52~(9) (2026)
  1107--1112.

\bibitem{JTM:2026}
J.~T. de~Miranda, M.~Khodas, A.~Levchenko, Magnetochiral anisotropy in strained
  superconducting transition metal dichalcogenides, Phys. Rev. B 114 (2026)
  154503.

\bibitem{HouzetMeyer:2015}
M.~Houzet, J.~S. Meyer, Quasiclassical theory of disordered {Rashba}
  superconductors, Phys. Rev. B 92 (2015) 014509.

\bibitem{VBT:2022}
P.~Virtanen, F.~S. Bergeret, I.~V. Tokatly, Nonlinear $\sigma$ model for
  disordered systems with intrinsic spin-orbit coupling, Phys. Rev. B 105
  (2022) 224517.

\bibitem{IlicBergeret:2024}
S.~Ili\ifmmode~\acute{c}\else \'{c}\fi{}, P.~Virtanen, D.~Crawford, T.~T.
  Heikkil\"a, F.~S. Bergeret, Superconducting diode effect in diffusive
  superconductors and {Josephson} junctions with {Rashba} spin-orbit coupling,
  Phys. Rev. B 110 (2024) L140501.

\bibitem{NunchotYanase:2026}
N.~Nunchot, Y.~Yanase, Superconducting diode effect in the weak localization
  regime, Phys. Rev. B 114 (2026) 024513.

\bibitem{FLS:2000}
M.~V. Feigel'man, A.~I. Larkin, M.~A. Skvortsov, Keldysh action for disordered
  superconductors, Phys. Rev. B 61 (2000) 12361.

\bibitem{KamenevAndreev:1999}
A.~Kamenev, A.~Andreev, Electron-electron interactions in disordered metals:
  Keldysh formalism, Phys. Rev. B 60 (1999) 2218.

\bibitem{KamenevBook}
A.~Kamenev, Field Theory of Non-Equilibrium Systems, Cambridge University
  Press, Cambridge, 2011.

\bibitem{LevchenkoKamenev:2007}
A.~Levchenko, A.~Kamenev, Keldysh {Ginzburg-Landau} action of fluctuating
  superconductors, Phys. Rev. B 76 (2007) 094518.

\bibitem{Liao:2017}
Y.~Liao, A.~Levchenko, M.~S. Foster, Response theory of the ergodic many-body
  delocalized phase: {Keldysh Finkel'stein} sigma models and the 10-fold way,
  Annals of Physics 386 (2017) 97--157.

\bibitem{BergeretTokatly:2014}
F.~S. Bergeret, I.~V. Tokatly, Spin-orbit coupling as a source of long-range
  triplet proximity effect in superconductor-ferromagnet hybrid structures,
  Phys. Rev. B 89 (2014) 134517.

\bibitem{VBT:2021}
P.~Virtanen, F.~S. Bergeret, I.~V. Tokatly, Magnetoelectric effects in
  superconductors due to spin-orbit scattering: Nonlinear $\sigma$-model
  description, Phys. Rev. B 104 (2021) 064515.

\bibitem{Kamenev:2009}
A.~Kamenev, A.~Levchenko, Keldysh technique and non-linear $\sigma$-model:
  basic principles and applications, Advances in Physics 58~(3) (2009)
  197--319.

\bibitem{KLB:1975}
R.~A. Klemm, A.~Luther, M.~R. Beasley, Theory of the upper critical field in
  layered superconductors, Phys. Rev. B 12 (1975) 877.

\bibitem{AL:1968}
L.~G. Aslamazov, A.~I. Larkin, Effect of fluctuations on the properties of a
  superconductor above the critical temperature, Fiz. Tverd. Tela 10 (1968)
  1104, [Sov. Phys. Solid State {\bf 10}, 875 (1968)].

\bibitem{LarkinVarlamov:2005}
A.~I. Larkin, A.~A. Varlamov, Theory of Fluctuations in Superconductors,
  Clarendon Press, Oxford, 2005.

\bibitem{Bagrets:2014}
D.~Bagrets, A.~Levchenko, Nonequilibrium noise and current fluctuations at the
  superconducting phase transition, Phys. Rev. B 90 (2014) 180505(R).

\bibitem{Forster:2008}
H.~F{\"o}rster, M.~B{\"u}ttiker, Fluctuation relations without
  microreversibility in nonlinear transport, Phys. Rev. Lett. 101 (2008)
  136805.

\bibitem{SaitoUtsumi:2008}
K.~Saito, Y.~Utsumi, Symmetry in full counting statistics, fluctuation theorem,
  and relations among nonlinear transport coefficients in the presence of a
  magnetic field, Phys. Rev. B 78 (2008) 115429.

\end{thebibliography}

\end{document}